\documentclass[12pt]{article}%
\usepackage[utf8]{inputenc}
\usepackage[T1]{fontenc}
\usepackage[nosort]{cite}
\usepackage{braket}
\usepackage{graphicx}
\usepackage{multicol}
\usepackage{amsfonts}
\usepackage{amssymb}
\usepackage{amsmath}
\usepackage{heck}
\usepackage{afterpage}
\usepackage{setspace}
\usepackage{verbatim}
\usepackage{color}
\usepackage{longtable}
\usepackage{float}
\usepackage{subcaption}
\usepackage{epsfig}
\usepackage{epstopdf}
\usepackage{adjustbox}
\usepackage{tikz}
\usepackage[margin=1in]{geometry}
\usepackage{titletoc}
\usepackage{hyperref}%
\usepackage{mathrsfs}

\providecommand{\U}[1]{\protect\rule{.1in}{.1in}}
\pdfoutput=1
\newsavebox{\mysavebox}

\hypersetup{colorlinks,citecolor=black,filecolor=black,linkcolor=black,urlcolor=black,pdftex}
\usetikzlibrary{decorations.markings}

\numberwithin{equation}{section}

\newcommand{\be}{\begin{equation}}
\newcommand{\ee}{\end{equation}}

\usepackage{tikz}
\usetikzlibrary{arrows}
\usetikzlibrary{arrows.meta}
\usetikzlibrary{arrows,decorations.pathmorphing}
\usetikzlibrary{shapes.geometric,calc,arrows, positioning,shapes.misc,decorations.markings}
\tikzset{
  big arrow/.style={
    decoration={markings,mark=at position 1 with {\arrow[scale=2,#1]{>}}},
    postaction={decorate},
    shorten >=0.4pt},
  big arrow/.default=black}

\pgfdeclarelayer{edgelayer}
\pgfdeclarelayer{nodelayer}
\pgfsetlayers{edgelayer,nodelayer,main}
\tikzstyle{none}=[inner sep=0pt]

\tikzstyle{NodeCross}=[draw, shape=circle, cross out, inner sep=0pt, minimum size=6pt,line width=0.25mm]
\tikzstyle{Circle}=[draw, shape=circle, black, fill=black, inner sep=0pt, minimum size=6pt]
\tikzstyle{circle}=[draw, shape=circle, black, fill=black, inner sep=0pt, minimum size=16pt]
\tikzstyle{Star}=[draw, shape=star, fill=red, star points=8, inner sep=0pt, minimum size=8pt]
\tikzstyle{CircleRed}=[draw, shape=circle, black, fill=red, inner sep=0pt, minimum size=6pt]
\tikzstyle{StarP}=[draw={rgb,255: red,128; green,0; blue,128}, shape=star, fill={rgb,256: red,128; green,0; blue,128}, star points=8, inner sep=0pt, minimum size=12pt]
\tikzstyle{ShadedCircRed}=[draw=red, shape=circle, fill={rgb, 255: red,255; green,114; blue, 118}, inner sep=0pt, minimum size=80pt, line width=0.5mm, fill opacity=0.2]
\tikzstyle{ShadedCircRed2}=[draw=red, shape=circle, fill={rgb, 255: red,255; green,114; blue, 118}, inner sep=0pt, minimum size=10pt]
\tikzstyle{ShadedCircRed3}=[draw=black, shape=rectangle, fill={rgb, 255: red,255; green,114; blue, 118}, inner sep=0pt, minimum size=113pt, line width=0.25mm]
\tikzstyle{ShadedCirc}=[draw=red, shape=circle, fill=white, inner sep=0pt, minimum size=45pt,  fill opacity=1.0,  line width=0.5mm]
\tikzstyle{CircleBlue}=[draw, shape=circle, fill=blue, inner sep=0pt, minimum size=6pt]
\tikzstyle{BigCirclePurple}=[draw, shape=circle, fill={rgb,255: red,191; green,0; blue,191}, inner sep=0pt, minimum size=12pt]
\tikzstyle{CirclePurple}=[draw, shape=circle, fill={rgb,255: red,191; green,0; blue,191}, inner sep=0pt, minimum size=8pt]
\tikzstyle{EmptyCircle}=[draw, shape=circle, inner sep=0pt, minimum size=4pt]
\tikzstyle{GreenCircle}=[draw, shape=circle,  fill={rgb,255: red,80; green,200; blue,120}, inner sep=0pt, minimum size=8pt]
\tikzstyle{BrownCircle}=[draw, shape=circle,  fill={rgb,255: red,210; green,105; blue,30}, inner sep=0pt, minimum size=8pt]
\tikzstyle{CirclePurpleSmall}=[draw, shape=circle, fill={rgb,255: red,191; green,0; blue,191}, inner sep=0pt, minimum size=4pt]
\tikzstyle{BigCircleGreen}=[draw, shape=circle, fill={rgb,255: red,0; green,191; blue,0}, inner sep=0pt, minimum size=12pt]
\tikzstyle{BigCircleBlue}=[draw, shape=circle, fill={rgb,255: red,0; green,0; blue,191}, inner sep=0pt, minimum size=12pt]
\tikzstyle{BigCircleRed}=[draw, shape=circle, fill={rgb,255: red,191; green,0; blue,0}, inner sep=0pt, minimum size=12pt]
\tikzstyle{BrownCircleSmall}=[draw, shape=circle,  fill={rgb,255: red,210; green,105; blue,30}, inner sep=0pt, minimum size=6pt]

\tikzstyle{DashedLine}=[-, densely dashed, line width=0.25mm]
\tikzstyle{DottedLine}=[-, dotted, line width=0.25mm]
\tikzstyle{ThickLine}=[-, line width=0.25mm]
\tikzstyle{ArrowLineRight}=[-, -{Stealth[scale=1.25]}, line width=0.25mm, scale=5]
\tikzstyle{ArrowLineRed}=[-, draw={rgb,255: red,191; green,0; blue,0}, -{Stealth[scale=1.75]}, line width=0.1mm, scale=5]
\tikzstyle{RedLine}=[-, draw={rgb,255: red,191; green,0; blue,0}, fill=none, line width=0.5mm]
\tikzstyle{DashedLineThin}=[-, densely dashed, line width=0.125mm, fill=none, draw=black]
\tikzstyle{DottedRed}=[-, dotted, draw={rgb,255: red,191; green,0; blue,0}, fill=none, line width=0.25mm]
\tikzstyle{DashedRed}=[-, densely dashed, draw={rgb,255: red,191; green,0; blue,0}, fill=none, line width=0.25mm]
\tikzstyle{BlueLine}=[-, draw={rgb,255: red,0; green,0; blue,191}, fill=none, line width=0.5mm]
\tikzstyle{ArrowLineBlue}=[-, draw={rgb,255: red,0; green,0; blue,191}, -{Stealth[scale=1.75]}, line width=0.1mm, scale=5]
\tikzstyle{GreenDoubleArrow}=[<->, draw={rgb,155: red,0; green,255; blue,0},  line width= 0.5mm, scale=5]
\tikzstyle{RedDoubleArrow}=[<->, draw={rgb,255: red,255; green,0; blue,0},  line width= 0.5mm, scale=5]
\tikzstyle{BlueDottedLight}=[-, dotted, draw={rgb,255: red,0; green,0; blue,191}, fill=none, line width=0.3mm]
\tikzstyle{BrownLine}=[-, draw={rgb,255: red,210; green,105; blue,30}, fill=none, line width=0.5mm]
\tikzstyle{DottedRed}=[-, dotted, draw={rgb,255: red,191; green,0; blue,0}, fill=none, dotted, line width=0.5mm]
\tikzstyle{DottedPurple}=[-, dotted, draw={rgb,255: red,191; green,0; blue,191}, fill=none, dotted, line width=0.5mm]
\tikzstyle{BlueDottedLight}=[-, dotted, draw={rgb,255: red,0; green,0; blue,191}, fill=none, line width=0.5mm]
\tikzstyle{ArrowLinePurple}=[-, draw={rgb,255: red,191; green,0; blue,191}, -{Stealth[scale=1.75]}, line width=0.5mm, scale=5]
\tikzstyle{DashedLineGreen}=[-, densely dashed, draw={rgb,255: red,74; green,103; blue,65}, line width=0.25mm]
\tikzstyle{LineGreen}=[-, draw={rgb,255: red, 74; green,200; blue,65}, line width=0.5mm]
\tikzstyle{ArrowLineGreen}=[-, draw={rgb,255: red,0; green,191; blue,0}, -{Stealth[scale=1.75]}, line width=0.5mm, scale=5]
\tikzstyle{GreenLine}=[-, draw={rgb,255: red,0; green,191; blue,0}, fill=none, line width=0.5mm]
\tikzstyle{PurpleLine}=[-, draw={rgb,255: red,191; green,0; blue,191}, fill=none, line width=0.5mm]
\tikzstyle{PPurpleLine}=[-, draw={rgb,255: red,191; green,0; blue,191}, fill=none, line width=2.5mm]
\tikzstyle{DPurpleLine}=[-, dotted, draw={rgb,255: red,191; green,0; blue,191}, fill=none, line width=0.5mm]
\tikzstyle{SBrownLine}=[-, draw={rgb,255: red,191; green,0; blue,191}, fill=none, opacity=0.35, line width=2.5mm]

\tikzset{snake it/.style={decorate, decoration=snake}}

\usetikzlibrary{decorations.markings}

\usetikzlibrary{hobby}

\tikzset{
dashstar/.style={
 dash pattern=on 5pt off 5pt,
 postaction={
  decorate,
  decoration={
   markings,
   mark=between positions 9pt and 1 step 10pt with {
     \node[color=red] {*};
   }
  }
 }
},
dashstarstar/.style={ 
 dash pattern=on 5pt off 10pt,
 postaction={
   decorate,
   decoration={
     markings,
     mark=between positions 10pt and 1
          step 15pt
           with {
            \node at (-2pt,0pt) {\pgfuseplotmark{star}};
            \node at (2pt,0pt) {\pgfuseplotmark{star}};
           }
   }
 }
}
}
\usepackage{pgfplots}
\pgfplotsset{compat=1.16}

\usetikzlibrary{cd}

\newcommand{\ba}{\begin{aligned}}
\newcommand{\ea}{\end{aligned}}

\newcommand\beq{\begin{equation}}
\newcommand\eeq{\end{equation}}

\begin{document}

\preprint{CERN-TH-2024-019\\ IFT-24-013}

\date{February 2024}

\title{On the Fate of \\[1mm] Stringy Non-Invertible Symmetries}

\institution{PENN}{\centerline{$^{1}$Department of Physics and Astronomy, University of Pennsylvania, Philadelphia, PA 19104, USA}}
\institution{CALTECH}{\centerline{$^{2}$Walter Burke Institute for Theoretical Physics, California Institute of Technology,
Pasadena, CA 91125, USA}}
\institution{UAMIFT}{\centerline{$^{3}$Instituto de F\'isica Te\'orica UAM-CSIC,
Universidad Aut\'onoma de Madrid, Cantoblanco, 28049 Madrid, Spain}}
\institution{SCGP}{\centerline{$^{4}$Simons Center for Geometry and Physics, SUNY, Stony Brook, NY 11794, USA}}
\institution{HARVARD}{\centerline{$^{5}$Department of Physics, Harvard University, Cambridge, MA 02138, USA}}
\institution{CERN}{\centerline{$^{6}$CERN, Theoretical Physics Department, 1211 Meyrin, Switzerland}}

\authors{Jonathan J. Heckman,\worksat{\PENN}\footnote{e-mail: \texttt{jheckman@sas.upenn.edu}}
Jacob McNamara,\worksat{\CALTECH}\footnote{email: \texttt{jmcnamar@caltech.edu}}
Miguel Montero,\worksat{\UAMIFT}\footnote{email: \texttt{miguel.montero@csic.es}} \\[4mm]
Adar Sharon,\worksat{\SCGP}\footnote{email: \texttt{asharon@scgp.stonybrook.edu}}
Cumrun Vafa,\worksat{\HARVARD}\footnote{e-mail: \texttt{vafa@g.harvard.edu}} and
Irene Valenzuela\worksat{\UAMIFT, \CERN}\footnote{e-mail: \texttt{irene.valenzuela@cern.ch}}
}

\abstract{\noindent
Non-invertible symmetries in quantum field theory (QFT) generalize the familiar product rule of groups to a more general fusion rule. In many cases, gauged versions of these symmetries can be regarded as dual descriptions of invertible gauge symmetries. One may ask: are there any other types of non-invertible gauge symmetries? In theories with gravity we find a new form of non-invertible gauge symmetry that emerges in the limit of fundamental, tensionless strings. These stringy non-invertible gauge symmetries appear in standard examples such as non-abelian orbifolds. Moving away from the tensionless limit always breaks these symmetries. We also find that both the conventional form of non-invertible gauge symmetries and these stringy generalizations are realized in AdS/CFT.
Although generically broken, approximate non-invertible symmetries have implications for Swampland constraints: in certain cases they can be used to prove the existence of towers of states related to the Distance Conjecture, and can sometimes explain the existence of slightly sub-extremal states which fill in the gaps in the sublattice Weak Gravity Conjecture.}

\maketitle

\tableofcontents

\newpage

\section{Introduction}

Symmetries play an important role in constraining the dynamics of quantum systems.
This is especially true in the case of an unbroken symmetry, where we can derive exact selection rules, but it also applies in situations where
the symmetry breaking is controlled by a small parameter.

Recently a number of investigations have suggested a generalization of global symmetries
in quantum field theory (QFT) beyond the more familiar group-like composition rule.
In this broader setting of \textit{non-invertible symmetry}, one defines a symmetry in terms of a topological operator that links with the
charged object of interest \cite{Gaiotto:2014kfa}. The product of two such topological operators
might end up realizing a more general fusion rule such as:
\begin{equation}
\mathcal{N}_i \otimes \mathcal{N}_j = \underset{k}{\sum} \mathcal{T}_{ij}^{k}\mathcal{N}_{k},
\end{equation}
where the $\mathcal{T}_{ij}^{k}$ denote c-number coefficients.\footnote{In general, $\mathcal{T}_{ij}^{k}$ is the partition function of a decoupled TQFT.} There are by now many examples of this sort, such as in 2D rational conformal field theories (CFTs) with non-invertible Verlinde lines \cite{Verlinde:1988sn}, in 4D gauge theories with a gauged charge conjugation symmetry \cite{Heidenreich:2021xpr}, as well as in many
other contexts.\footnote{The literature has substantially grown in the past few years. For reasonably up to date reviews, see e.g., the reviews \cite{Cordova:2022ruw, Schafer-Nameki:2023jdn, Bhardwaj:2023kri, Luo:2023ive, Brennan:2023mmt, Shao:2023gho} and references therein.}

What becomes of these non-invertible symmetries in quantum gravity? On general grounds one expects that unbroken symmetries are ``gauged,'' namely they
instead specify a redundancy in physical configurations.
In the context of QFT with gravity switched off, there is a notion of ``gauging a non-invertible symmetry'' by inserting a mesh of topological operators (see e.g., \cite{Fuchs:2002cm, Carqueville:2012dk, Bhardwaj:2017xup}).
We can view this as producing a notion of non-invertible gauge theory.\footnote{See also \cite{kawagoe2024levinwen} for a recent discussion of the sense in which this procedure can really be viewed as producing a ``gauge theory.''} Based on this, it is natural to ask whether we can produce examples of this sort of non-invertible symmetries
directly in quantum gravity. Here we find a few surprises, both from the point of view of worldsheet constructions, and also from the perspective of the AdS/CFT correspondence.

To begin, recall that global symmetries of a string worldsheet theory correspond to gauge symmetries in the target space \cite{Banks:1988yz}. From this perspective, it would seem that all we require is an example of a non-invertible global symmetry in the 2D worldsheet CFT. Such non-invertible worldsheet symmetries were recently discussed in the context of string theory in \cite{Cordova:2023qei}. As a simple example, consider an orbifold as specified by a non-abelian group $\Gamma$; the operation of gauging a global $\Gamma$ symmetry
on the worldsheet results in an orbifold theory with Hilbert space sectors labelled by conjugacy classes of $\Gamma$ (see \cite{Dixon:1985jw, Dixon:1986jc}). In categorical terms, gauging the $\Gamma$ symmetry results in a ``magnetic'' global zero-form categorical symmetry $\mathrm{Rep} (\Gamma)$.\footnote{See references \cite{Bhardwaj:2017xup, Tachikawa:2017gyf}, and for a complementary perspective see e.g., reference \cite{Robbins:2021ibx, Sharpe:2021srf}.} In this case, symmetry operators are labelled by representations of $\Gamma$ and we get a nontrivial fusion rule whenever $\Gamma$ is non-abelian (more than one summand can appear in the fusion of two irreducible representations of $\Gamma$). So, the appearance of this non-invertible \textit{global} symmetry in the 2D CFT would seem to suggest the existence of a corresponding non-invertible \textit{gauge} symmetry in the target space.

But string theory is more than just a 2D CFT; it also involves coupling this system to 2D gravity! Moreover, it is well-known that the selection rules arising from $\mathrm{Rep} (\Gamma)$ symmetry are violated at higher-loop order \cite{Hamidi:1986vh, Dijkgraaf:1989hb}, leaving only the selection rules corresponding to representations $\mathrm{Rep}(\Gamma_{\mathrm{ab}})$ of the abelianization, an invertible symmetry. So, while tree level string theory appears to enjoy a non-invertible $\mathrm{Rep} (\Gamma)$ symmetry, it is broken by $g_{s} \neq 0$ effects. This is an example of a more general phenomenon \cite{McNamara:2021cuo}: when a QFT in any dimension is coupled to semiclassical gravity, the appearance of nontrivial topologies in the gravitational path integral leads to a generic breaking of all non-invertible symmetries in the absence of extreme cancellations. We also present a similar construction of non-invertible symmetries broken at $g_s\neq 0$ for the case of string theory on toroidal orbifolds.

The lesson we draw is that the target space physics of non-invertible worldsheet symmetries does not correspond to the QFT notion of non-invertible gauge symmetry coming from ``summing over a mesh of topological operators''. Indeed, to see the non-invertible worldsheet symmetry emerge we must take $g_s \to 0$ and consider the entire structure of
perturbative string theory, taking us well outside the regime of local effective field theory. We will describe the target space physics of non-invertible worldsheet symmetries as ``stringy non-invertible gauge symmetry,'' which is generically Higgsed, but which can be restored in a limit where the string becomes tensionless in Planck units.

Given this state of affairs, it is natural to ask whether the breaking of this sort of non-invertible gauge symmetry at $g_s \neq 0$ is merely an artifact of this specific class of examples, or is something that holds more generally in quantum gravity. Along these lines, we consider another context where non-invertible gauge symmetries seem easy to realize in quantum gravity: examples from holography in which the CFT of an AdS/CFT pair enjoys a non-invertible symmetry. Indeed, recently there has been progress in realizing examples of non-invertible symmetries in a number of stringy and holographic constructions \cite{Apruzzi:2022rei,
GarciaEtxebarria:2022vzq, Heckman:2022muc, Heckman:2022xgu, Dierigl:2023jdp,
Cvetic:2023plv, Bah:2023ymy, Apruzzi:2023uma, Cvetic:2023pgm, Heckman:2024oot,Antinucci:2022vyk}. From this perspective, a non-invertible \textit{global} symmetry
of the boundary theory would seem to automatically imply the existence of a \textit{gauged} non-invertible symmetry in the bulk.

In all examples with a semiclassical bulk, this is indeed the case, as has previously been discussed in the literature: the bulk contains a topological sector described by the symmetry topological field theory (SymTFT) (see e.g., \cite{Reshetikhin:1991tc, Turaev:1992hq, Barrett:1993ab, Witten:1998wy, Fuchs:2002cm, Kirillov:2010nh, Kapustin:2010if, Kitaev:2011dxc, Fuchs:2012dt,
Freed:2012bs, Freed:2018cec, Apruzzi:2021nmk, Freed:2022qnc, Kaidi:2022cpf,
Baume:2023kkf,Brennan:2024fgj, Antinucci:2024zjp}). For a non-invertible global symmetry on the boundary, the corresponding SymTFT can be understood as a non-invertible bulk gauging of the boundary symmetry in the conventional sense.\footnote{This is always true in the sense of ``summing over a mesh of topological operators,'' which are given by condensates of the gapped boundary condition (see e.g., \cite{Fuchs:2002cm, Carqueville:2012dk, Bhardwaj:2017xup}). We expect further that the very recent work \cite{kawagoe2024levinwen} for 3D SymTFTs generalizes to any dimension, so any SymTFT can be viewed as a gauge theory for the higher tube algebra \cite{Bartsch:2022mpm, Bartsch:2022ytj, Bartsch:2023wvv, Bhardwaj:2023wzd, Bhardwaj:2023ayw} of the boundary symmetry.} However, it is also worth noting that in every example we consider, the sense in which the gauge symmetry in the bulk is non-invertible is rather benign: this topological sector admits a more conventional characterization as an invertible gauge theory, possibly after switching to a dual basis of fields. If the bulk quantum gravity theory admits a worldsheet description, the invertible gauge theory presentation is more natural, as it is the one that acts most naturally on the string worldsheet.

A particularly instructive example in this regard is the background AdS$_3 \times S^3 \times T^4$. For tuned values of the $T^4$ moduli, we can have a non-abelian symmetry $\Gamma$ acting on $T^4$, leading to a $\Gamma$ gauge theory propagating in the AdS$_3 \times S^3$ factor. For suitable boundary conditions we get a $\mathrm{Rep} (\Gamma)$ symmetry in the boundary CFT$_2$, and the bulk $\Gamma$ gauge theory in 6D could dually be viewed as a 3-form $\mathrm{Rep}(\Gamma)$ gauge theory (reducing on the $S^3$ factor, this is a $0$-form $\mathrm{Rep}(\Gamma)$ gauge theory in AdS$_3$). Nevertheless, if we consider the worldsheet description of the bulk (or at least the $T^4$ factor), the symmetry that acts on the worldsheet is still just $\Gamma$.

Moving beyond semiclassical bulks, we can also consider the limit of AdS$_3 \times S^3 \times T^4$ with only a single unit of flux, such that the bulk is described by a tensionless string theory \cite{Gaberdiel:2018rqv, Eberhardt:2018ouy, Eberhardt:2019ywk}. In this limit, we expect an enormous non-invertible symmetry to emerge in the CFT: the CFT dual is given by the symmetric orbifold CFT $\mathrm{Sym}^N(T^4)$.\footnote{If we choose boundary conditions that realize $\mathrm{Rep}(\Gamma)$ symmetry, we should strictly speaking discuss the $\Gamma$ orbifold $\mathrm{Sym}^N(T^4)/\Gamma$.} Exactly at the symmetric orbifold point, this CFT$_2$ admits a non-invertible $\mathrm{Rep}(S_N)$ symmetry, where $S_N$ is the symmetric group. We find that the bulk dual of this is not any conventional non-invertible gauge symmetry in AdS$_3 \times S^3 \times T^4$. Instead, the large $N$ limit $\mathrm{Rep}(S_\infty)$ is realized as a non-invertible global symmetry of the tensionless string theory worldsheet, which remains unbroken as a result of the extreme cancellations that appear in the tensionless limit. We do not have a general characterization of which non-invertible symmetries of holographic CFTs are realized as conventional non-invertible bulk gauge symmetries or as stringy non-invertible gauge symmetries, but it is worth noting that the the quantum dimensions of topological operators for $\mathrm{Rep}(S_N)$ symmetry scale with $N$, while the $\mathrm{Rep}(\Gamma)$ symmetry does not.\footnote{It is not enough to have the size of the symmetry scale with $N$: for example, even the invertible $\mathbb{Z}_N$ center 1-form symmetry of $\mathcal{N} = 4$ $SU(N)$ super Yang-Mills scales with $N$. Note also that in orbifolds based on the $D_N$-series of finite subgroups in $SU(2)$, the order of the group can be parametrically large, but the dimensions of irreducible
representations remains small so that the fusion rule is still rather benign.}

The common theme in these examples is that in quantum gravity, less benign forms of gauged non-invertible symmetries are generically broken and only seem  to emerge in special limits in field space (like the tensionless string limit $g_s\rightarrow 0$). This implies that the breaking effects become suppressed in these limits, so they appear as \emph{approximate symmetries} in the effective field theory. Therefore, despite being broken, they can still have interesting applications in the string landscape. In particular,
they fit rather well with a number of related Swampland considerations connected with infinite distance limits (see \cite{Brennan:2017rbf,Palti:2019pca,vanBeest:2021lhn,Agmon:2022thq} for reviews). The most important result is that the presence of a non-invertible symmetry in the worldsheet combined with modular invariance implies the existence of an infinite tower of states which is charged under the non-invertible symmetry and becomes light at infinite distance, as predicted by the Distance Conjecture.  In these cases this allows us to generalize the usual worldsheet proof of the Weak Gravity Conjecture \cite{Heidenreich:2016aqi,Montero:2016tif,Heidenreich:2024dmr} to the case in which the tower is not charged under a massless gauge field. We will also see that in these cases the approximate non-invertible symmetries provide a complementary perspective on a number of subtle examples for several swampland conjectures. In particular, they can sometimes explain the existence of slightly sub-extremal states which fill in the gaps in the Sublattice Weak Gravity Conjecture. Moreover, in certain cases, the existence of 4D $\mathcal{N} = 2$ theories with properties more analogous to theories with $\mathcal{N} = 4$ supersymmetry can be reinterpreted in terms of the existence of a non-invertible symmetry.   Even though these features also have other more general explanations not related to non-invertible symmetries, it is satisfying to find a complementary explanation of these connected to symmetry principles, at least in certain cases.

The rest of this paper is organized as follows. In section \ref{sec:2DEXAMPLES} we review the fact that non-invertible symmetries of the worldsheet CFT are generically broken by string loop effects, and illustrate this effect in a few concrete examples. In section \ref{sec:ADS} we discuss examples in AdS/CFT. In section \ref{sec:SWAMP} we prove the existence of an infinite tower of states charged under the non-invertible symmetry that becomes light at infinite field distance. We then explain the different applications of the weakly broken non-invertible symmetry for several swampland considerations. We present a broader discussion and some avenues for future investigation in section \ref{sec:FUTURE}.

\textbf{Note Added:} As we were completing this work, we learned of \cite{Kaidi:2024} which we understand will also discuss non-invertible worldsheet symmetries.

\section{Non-Invertible Symmetry Breaking by String Loops} \label{sec:2DEXAMPLES}

In this section, we argue that any non-invertible symmetries of the worldsheet CFT are generically broken by string loop effects, in the absence of conspiracies. More concretely, we review the well-known fact that the selection rules placed on sphere correlation functions by non-invertible symmetry fail to hold at higher genus. A simple example is the energy operator $\varepsilon$ of the 2D Ising model: while $\varepsilon$ is charged under the non-invertible Kramers-Wannier symmetry, it picks up a nonzero one-point function on Riemann surfaces of positive genus \cite{Itzykson:1986pj, DiFrancesco:1987ez, Bagger:1988yc, Behera:1989gg}.

As a result of this breaking effect, if we use string perturbation theory to compute some scattering process forbidden at tree level by a non-invertible symmetry, we will generically pick up nonzero contributions at higher order in the string coupling $g_s$. From the perspective of target space physics, this means that the non-invertible gauge symmetry is only visible as an approximate symmetry for $g_s$ small. In contrast, any invertible symmetry of the worldsheet is preserved to all orders in $g_s$. Note that in Planck units, the limit $g_s \to 0$ in flat space corresponds to the tensionless limit of the string.

This section is organized as follows. First, we briefly review the general form of tree-level selection rules imposed by non-invertible symmetry derived in \cite{Lin:2022dhv}, and explain why these selection rules can fail to hold at higher genus. We then illustrate this effect in examples of non-invertible symmetries in familiar string compactifications. Finally, we comment on the story for a general non-invertible symmetry of a 2D CFT.

\subsection{Selection Rules from Topological Operators}\label{sec:SELECTIONRULES}

As stated in the Introduction, the modern understanding of symmetries in QFT is based around the notion of topological extended operators. In this section, we will focus on 0-form symmetries of 2D CFTs, which are generated by topological defect lines (TDLs). For a comprehensive discussion of TDLs in 2D CFTs, see \cite{Chang:2018iay}.

Let us recall the standard derivation of selection rules for an invertible symmetry using the associated TDL $\mathcal{U}$. Consider a sphere correlation function $\langle \mathcal{O}_1 \cdots \mathcal{O}_n \rangle$ of local operators transforming as $\mathcal{O}_i \to e^{i q_i} \mathcal{O}_i$ under the action of $\mathcal{U}$. We can nucleate a small loop of $\mathcal{U}$, pass it through the various operators and then annihilate it ``at infinity,'' leading to the same correlation function weighted by the sum of the charges:
\begin{equation}
    \langle \mathcal{O}_1 \cdots \mathcal{O}_n \rangle = e^{i \sum_i q_i} \langle \mathcal{O}_1 \cdots \mathcal{O}_n \rangle
\end{equation}
If the sum of charges is nonzero, the correlator must vanish, and so we have derived a selection rule from the presence of an invertible symmetry.

\begin{figure}[t!]
\begin{center}
\includegraphics[scale = 0.75, trim = {2.0cm 6.0cm 1.0cm 6.0cm}]{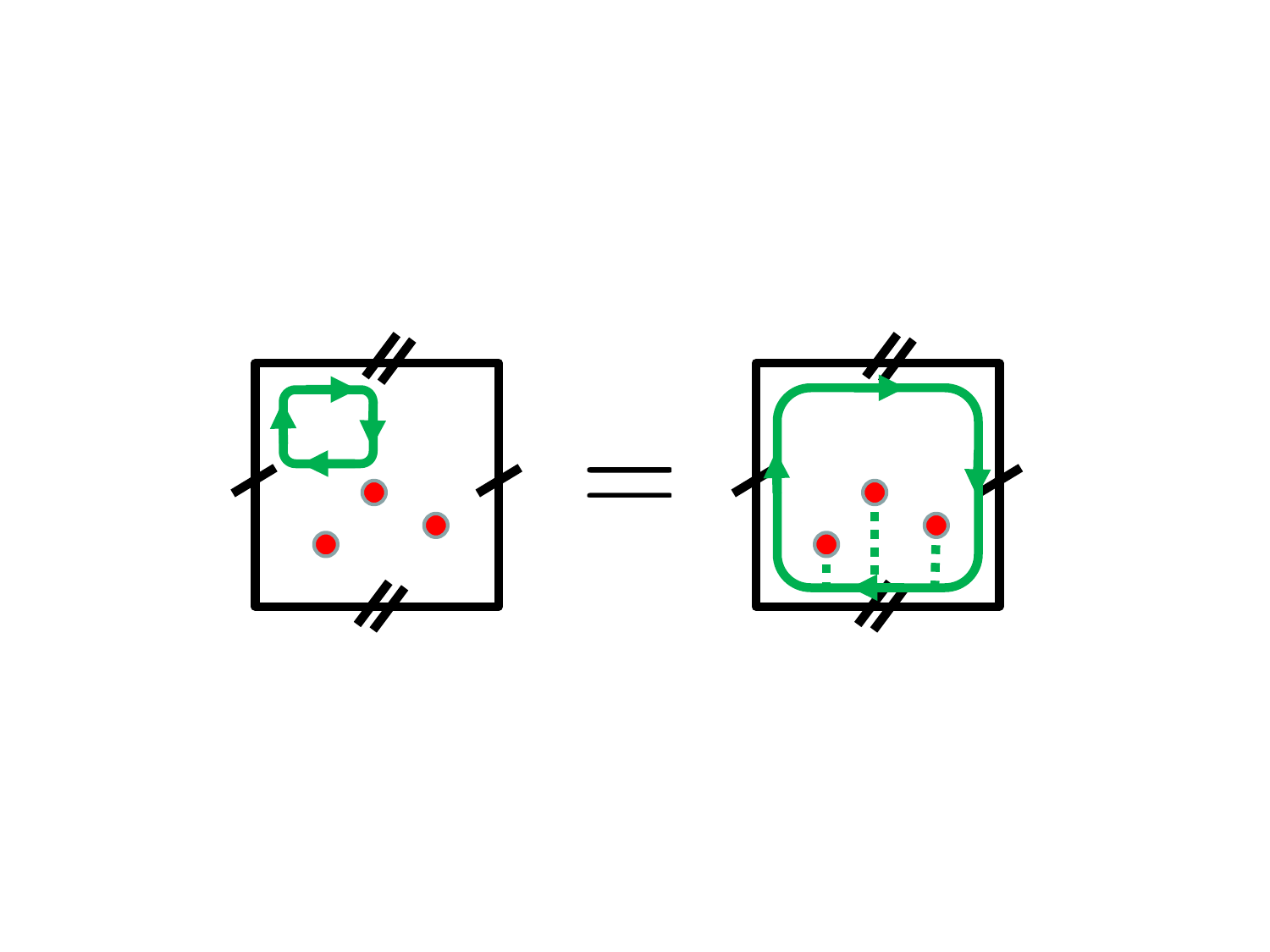}
\caption{Attempting to derive selection rules for a non-invertible symmetry on torus correlators.
Starting with our correlator, we nucleate a topological line (green), and pass it through the various local operators (red) picking up weights corresponding to their charges. Since the operator is non-invertible, there may be a network of lines (green, dashed) which attach the operators back to the topological operator (see \cite{Chang:2018iay} for a comprehensive discussion). Moreover, once we have pushed our topological line past all the local operator, we still cannot annihilate it, and instead are left with the fusion $\mathcal{N} \otimes \mathcal{N}^\dagger$ wrapped on the two nontrivial cycles of $\mathbb{T}^2$.}
\label{fig:selection_rules}
\end{center}
\end{figure}

What would happen if we tried to perform the same argument for a non-invertible TDL $\mathcal{N}$? First of all, when we nucleate a loop of $\mathcal{N}$, we would pick up a factor of the quantum dimension $\langle\mathcal{N}\rangle$; however, this factor will cancel when we annihilate $\mathcal{N}$ at infinity, so we ignore it. More interestingly, as we sweep $\mathcal{N}$ past any local operators, we might leave behind some network of TDLs, producing a correlation function of both local operators and disorder operators, i.e., point operators attached to topological lines (see Figure \ref{fig:selection_rules}). If we do produce such a network, then rather than deriving a constraint on a single correlation function, we might instead derive relationships between different correlation functions.

In fact, we could have run into a similar issue when deriving selection rules for invertible symmetries if we had not chosen our local operators to have definite charge, especially if our symmetry group were non-abelian. In the invertible context, the solution is well known: we should organize our operators $\mathcal{O}_i$ into representations $\mu_i$ of our symmetry group. A correlation function can only be nonzero provided the fusion $\mu_1 \otimes \cdots \otimes \mu_n$ of the representations includes a copy of the trivial representation.

This motivates us to organize our operators into ``representations'' of the action of the non-invertible symmetry. Importantly, a given ``representation'' might involve both local operators and disorder operators. In general, if our TDLs form a fusion category $\mathcal{C}$, these charges for our non-invertible symmetry are given by representations of Ocneanu's tube algebra $\mathrm{Tube}(\mathcal{C})$ (see e.g., \cite{OcneanuTube}), or equivalently \cite{izumi2000structure, muger2003subfactors}, by objects $\mu_i$ in the Drinfeld center $\mathcal{Z}(\mathcal{C})$. The most general selection rule for non-invertible symmetry tells us that a sphere correlation function involving local operators and disorder operators can only be nonzero if the fusion $\mu_1 \otimes \cdots \otimes \mu_n$ includes the trivial representation \cite{Lin:2022dhv}. While this abstract characterization is very powerful, we will not use it directly in examples below, and instead describe selection rules on sphere correlation functions on a case-by-case basis.

What goes wrong with the argument when we consider correlation functions on a more general Riemann surface $\Sigma$, such as the torus (see Figure \ref{fig:selection_rules})? Locally, we can proceed as before: we nucleate a loop of our non-invertible TDL $\mathcal{N}$ and pass it through our operators, possibly leaving behind a network of TDLs as before. The issue appears in the final step, where we attempt to annihilate $\mathcal{N}$ ``at infinity.'' In addition to possibly getting caught on our local operators in the correlation function, $\mathcal{N}$ may also get caught on the nontrivial topology of our Riemann surface $\Sigma$. Thus, in addition to the network of TDLs connecting our local operators, we pick up a network of the fusion $\mathcal{N} \otimes \mathcal{N}^\dagger$ wrapping the noncontractible cycles of $\Sigma$.\footnote{To make this argument precise for general $\Sigma$, choose a Morse function on $\Sigma$, and sweep $\mathcal{N}$ down $\Sigma$ according to the level sets of the Morse function. Each time we pass a Morse critical point of index one, $\mathcal{N}$ will get caught, leaving behind an insertion of the fusion $\mathcal{N} \otimes \mathcal{N}^\dagger$ on the descending manifold. This argument can be generalized to a non-invertible 0-form symmetry in any number of dimensions, where we will leave behind condensates built from $\mathcal{N}$ of various dimension on the descending manifolds of each Morse critical point of index $0 < i < n$.} Note that the fusion $\mathcal{N} \otimes \mathcal{N}^\dagger$ is the identity operator if and only if $\mathcal{N}$ is invertible!

We will return to the meaning of this particular network of TDLs in section \ref{sec:FUTURE}. For now, let us note that the appearance of an additional network of TDLs spoils the derivation of selection rules, so that whatever selection rules hold for sphere correlation functions need not hold on a general Riemann surface $\Sigma$. If we were simply studying 2D CFT, then this effect could be viewed as a mixed gravitational anomaly of any non-invertible symmetry:\footnote{Not to be confused with two distinct notions of anomaly for non-invertible symmetry that have previously been considered. In the case of invertible symmetries, one can consider three equivalent notions: the obstruction to gauging, the obstruction to a trivially gapped phase, or the violation of Ward identities in the presence of a background field. In the case of non-invertible symmetry these notions are different, and the one we mean is the third, where we view the manifold on which we place our CFT as a background gravitational field. For further discussion see
references \cite{Jensen:2017eof, Thorngren:2020yht}.} the selection rules that hold on the sphere are violated in the presence of a background topology. However, in the context of string theory, this violation of selection rules constitutes a genuine breaking of the symmetry, since we have made the worldsheet topology dynamical.

\subsection{Example: Non-Abelian Orbifolds}\label{sec:NONABELIAN}

Our first class of examples of non-invertible symmetry in the worldsheet CFT are provided by non-abelian orbifolds \cite{Dixon:1985jw, Dixon:1986jc}. Suppose we have a 2D CFT with non-abelian symmetry group $\Gamma$, and we form the orbifold CFT by gauging $\Gamma$. The twisted sectors are labeled by conjugacy classes $[g] \subset \Gamma$, whose fusion is defined as follows \cite{Dijkgraaf:1989hb}: given conjugacy classes $[g], [h]$, choose representative elements $g \in [g], h \in [h]$,\footnote{This is a slight abuse of notation, where we use the same symbol $g$ to denote the different possible representatives of its orbit $[g]$ under conjugation. We will continue to use this abuse of notation for representatives of orbits throughout this section for the purpose of readability.} and form the conjugacy class of their product $[g h]$. The fusion $[g] \otimes [h]$ is the sum of all conjugacy classes produced this way for different choices of $g, h$ modulo simultaneous conjugation by $\Gamma$.

From this description, we can easily derive selection rules on sphere correlation functions of twisted sector operators. If we have a sphere correlation function
\begin{equation}\label{eq:twisted_correlator}
    \langle \mathcal{O}_{[g_1]} \cdots \mathcal{O}_{[g_n]} \rangle_{\mathbb{S}^2}
\end{equation}
of operators in twisted sectors $[g_i]$, this correlation function can only be nonzero if the fusion $[g_1] \otimes \cdots \otimes [g_n]$ contains the conjugacy class $[1]$ of the identity element. But this is true if and only if we can choose some representatives $g_i \in [g_i]$ such that the product $g_1 \dots g_n = 1$ in the group $\Gamma$.

Let us now re-derive this selection rule using non-invertible topological operators (this is essentially the argument in \cite{Hamidi:1986vh}). In the orbifold theory, we have topological Wilson line operators $\mathcal{W}_\rho$ labeled by representations $\rho \in \mathrm{Rep}(\Gamma)$, which fuse according to the fusion of representations. Thus, the orbifold theory has $\mathrm{Rep}(\Gamma)$ non-invertible symmetry. Given a sphere correlation function \eqref{eq:twisted_correlator} of twisted sector operators as before, for any representation $\rho$, we can insert $\mathcal{W}_\rho$ along a loop encircling each of the operator insertions in turn. By annihilating this insertion ``at infinity,'' we learn that the holonomy around the loop must act trivially in $\rho$. Since $\rho$ was arbitrary, we learn that the holonomy must be the identity element $1 \in \Gamma$. But this holonomy is also the product in $\Gamma$ of the holonomies around each twisted sector operator, given by some representative elements $g_i \in [g_i]$ that multiply to the identity, so we have recovered the selection rule.

\begin{figure}[t!]
\begin{center}
\includegraphics[scale = 0.5, trim = {0.0cm 1.5cm 0.0cm 1.5cm}]{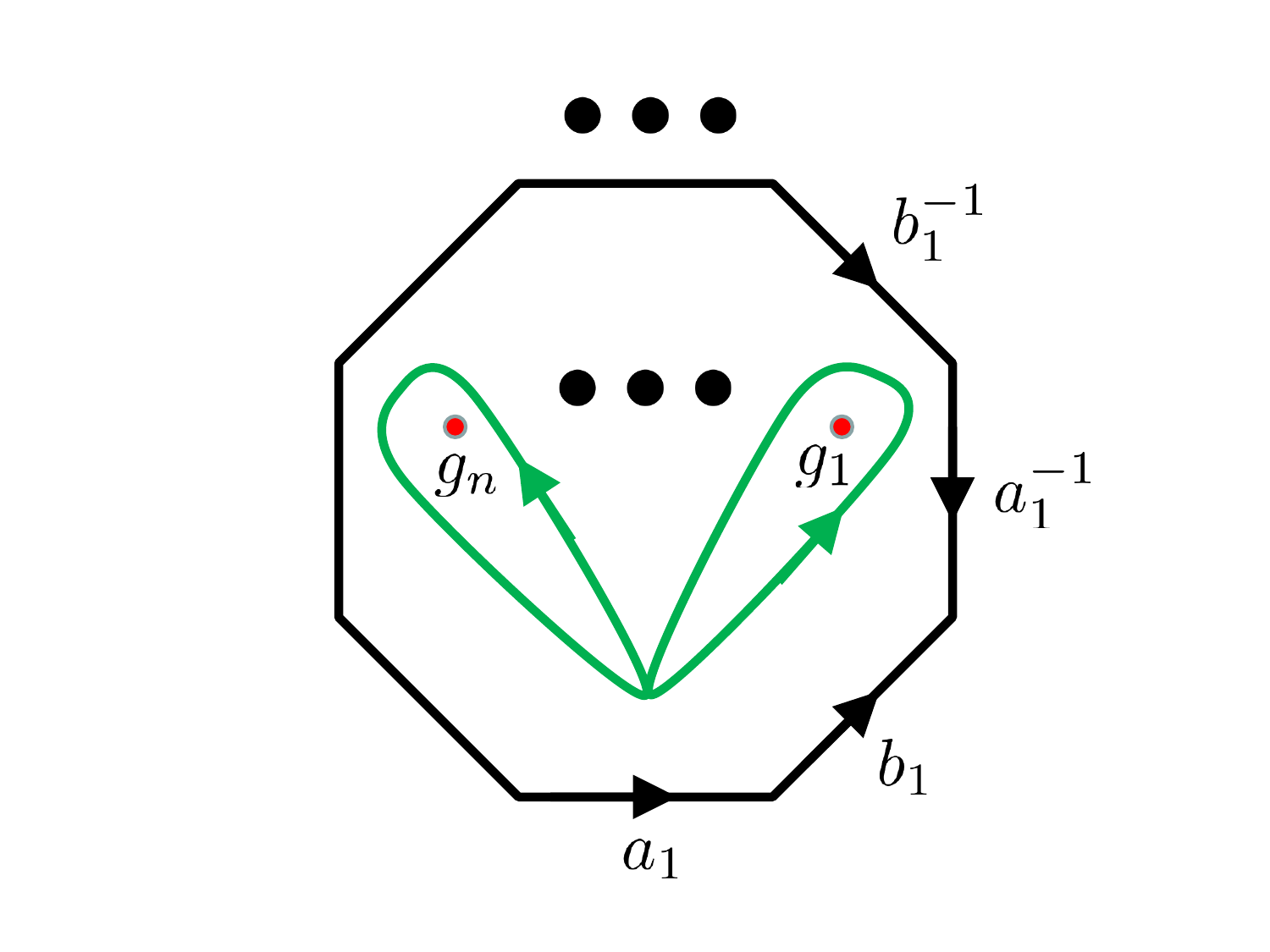}
\caption{We can construct a Riemann surface of genus $g$ by gluing the edges of a $4g$-gon in the pattern specified in \eqref{eq:product_of_commutators}. When we encircle a collection of twisted-sector operator insertions (red) with a topological Wilson line (green), we conclude that the holonomy around our collection of operators is a product of $g$ commutators in $\Gamma$. See also \cite[Figure 11]{Hamidi:1986vh}.}
\label{fig:trivial_holonomy}
\end{center}
\end{figure}

What happens on a nontrivial Riemann surface $\Sigma$ of genus $g$?\footnote{By another abuse of notation we shall use the same letter $g$ to now refer also to a genus. It should be clear from the context which notion is meant.} Following \cite{Hamidi:1986vh}, we note that a Riemann surface of genus $g$
can be formed by gluing the edges of a $4g$-gon in the pattern
\begin{equation}\label{eq:product_of_commutators}
    a_1 b_1 a_1^{-1} b_1^{-1} \dots a_g b_g a_g^{-1} b_g^{-1}
\end{equation}
as depicted in Figure \ref{fig:trivial_holonomy}. Now, when we push the topological Wilson line $\mathcal{W}_\rho$ ``to infinity,'' we cannot simply annihilate $\mathcal{W}_\rho$. Instead, we pick up the action in $\rho$ of an element in the commutator subgroup $[\Gamma, \Gamma] \subset \Gamma$ formed from a product of $g$ commutators $g h g^{-1} h^{-1}$. Thus, for a correlation function
\begin{equation}\label{eq:twisted_correlator_higher_genus}
    \langle \mathcal{O}_{[g_1]} \cdots \mathcal{O}_{[g_n]} \rangle_{\Sigma}
\end{equation}
of twisted sector operators to be nonzero, it is enough for the fusion $[g_1] \otimes \cdots \otimes [g_n]$ to contain the conjugacy class of a product of $g$ commutators in $G$. Note that the conjugacy classes of commutators $[g h g^{-1} h^{-1}]$ are precisely those that appear in fusions $[g] \otimes [g^{-1}]$ of conjugacy classes with their inverses. See e.g., \cite{McNamara:2021cuo} and references therein for further discussion.

What selection rules are preserved on all Riemann surfaces? In other words, how can we tell if a fusion product $[g_1] \otimes \cdots \otimes [g_n]$ does not contain a conjugacy class in the commutator subgroup $[\Gamma, \Gamma]$, so that \eqref{eq:twisted_correlator_higher_genus} must vanish at any genus? The answer is straightforward: the fusion product $[g_1] \otimes \cdots \otimes [g_n]$ lands in the commutator subgroup if and only if images of $[g_i]$ in the abelianization $\Gamma_{\mathrm{ab}} = \Gamma/[\Gamma, \Gamma]$ multiply to the identity. In other words, the conjugacy class $[g_i]$ carries a charge valued in $\Gamma_{\rm ab}$ given by its image, and these charges must cancel on any Riemann surface.

These charges are, in fact, simply the charges of the operators $\mathcal{O}_i$ under an invertible symmetry \cite{McNamara:2021cuo}. While the $\mathrm{Rep}(\Gamma)$ symmetry is, in general, non-invertible, it contains an invertible sub-symmetry, generated by invertible Wilson lines $\mathcal{W}_\rho$ corresponding to one-dimensional representations $\rho$. These invertible Wilson lines can still be ``annihilated at infinity'' even on a nontrivial Riemann surface, and so they impose the same selection rules on every Riemann surface. Moreover, the commutator subgroup $[\Gamma, \Gamma]$ must act trivially in any one-dimensional representation, and we have that the set of one-dimensional Wilson lines forms a $\mathrm{Rep}(\Gamma_{\rm ab}) = \Gamma_{\rm ab}^\vee$ invertible sub-symmetry of our non-invertible $\mathrm{Rep}(\Gamma)$ symmetry, which is the maximal invertible sub-symmetry.

This example illustrates an important point: while we expect any non-invertible symmetry to be broken down to its maximal invertible sub-symmetry by string loop effects, it is not true that the consequences of this breaking are always entirely visible at one loop. For example, suppose $\Gamma$ were a group such that some element $g_0$ in the commutator subgroup could only be written as a product of commutators, but not as a single commutator (such $\Gamma$ exist, see e.g. \cite{isaacs1977commutators} for a source of examples). Then a local operator in the twisted sector $[g_0]$ could not have a nonzero torus partition function, but could have a nonzero partition function on some higher-genus Riemann surface.

\begin{figure}[t!]
\begin{center}
\includegraphics[scale = 0.5, trim = {1.5cm 2.0cm 0.0cm 2.0cm}]{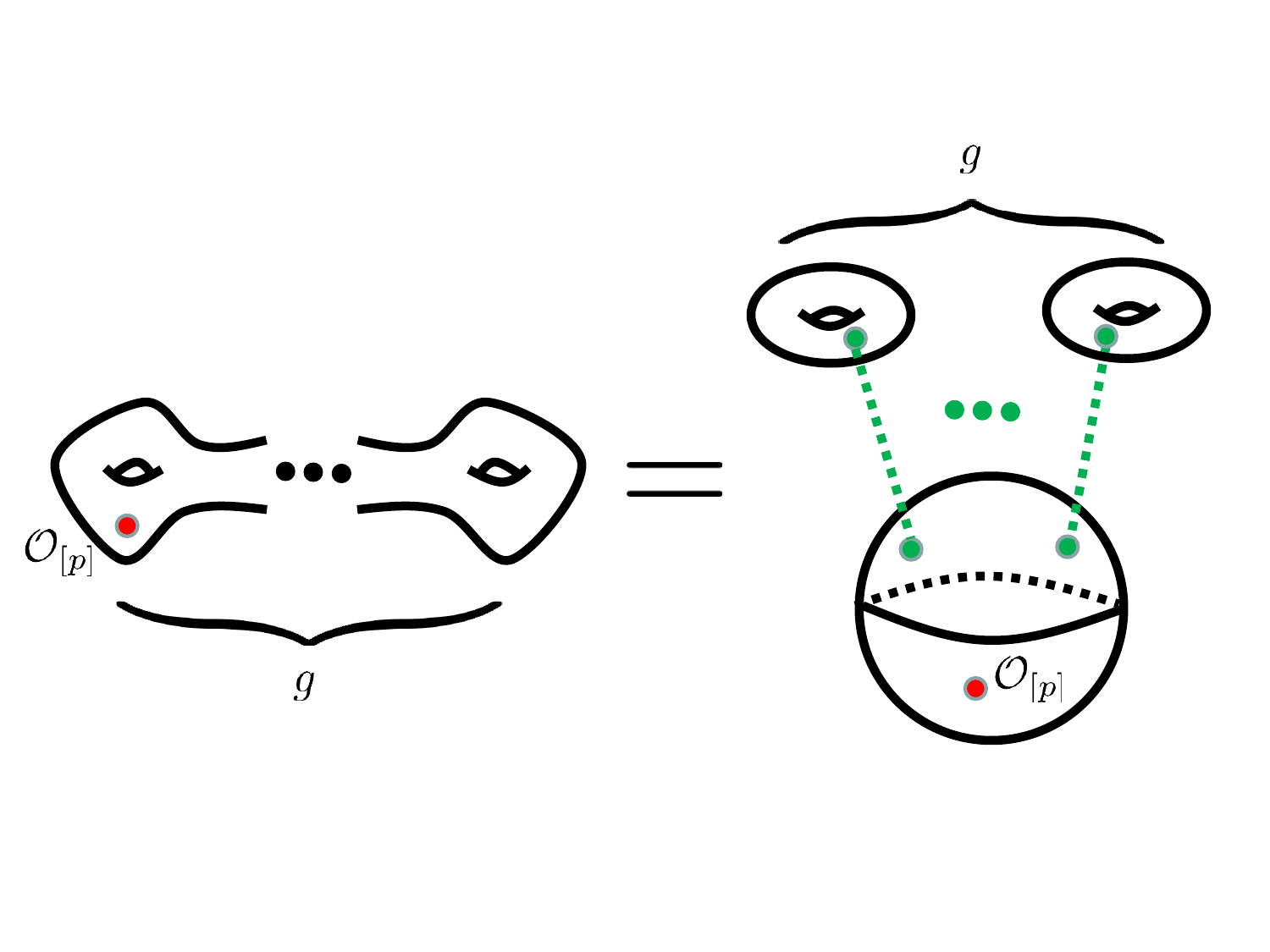}
\caption{A one-point function on a Riemann surface of genus $g$ (left) can be built be sewing together $g$ torus one-point functions and one sphere $(g+1)$-point function (right).}
\label{fig:higher_genus}
\end{center}
\end{figure}

However, in general, the set of charges that can get a nonzero one-point function at some order in the string loop expansion is always generated under fusion by the charges that can get a nonzero torus one-point function. This can be seen by realizing that a one-point function on a Riemann surface of genus $g$ can be built by sewing together $g$ torus one-point functions with a single sphere $(g+1)$-point function (see Figure \ref{fig:higher_genus}). Each torus one-point function produces some charge, and these charges simply fuse in the sphere $(g+1)$-point function. Thus, whatever symmetry is broken by string loops must be entirely broken at one-loop, even if the nonzero one-point functions of certain charges do not show up until higher loop order.

\subsection{Example: Toroidal Orbifolds}\label{sec:continuous_noninvertible}

Our next class of examples of non-invertible symmetry in the worldsheet CFT are given by toroidal orbifolds that break some part of the translation symmetry. These examples fall into the general class of non-invertible symmetries obtained by gauging a non-normal subgroup of a larger symmetry group (see e.g., \cite{Nguyen:2021yld, Heidenreich:2021xpr, Bhardwaj:2022yxj}); similar statements could be made for any of these more general examples, and in fact we will discuss one such generalization below in section \ref{sec:SUSY}.
These non-invertible symmetries capture, in the case of toroidal orbifolds, the general perturbative string theory expectation that tree-level scattering amplitudes are independent of the choice of compactification for states whose existence is unchanged by the compactification.\footnote{For example the tree-level scattering of graviton amplitudes in 4D is the same as those in 10D at string tree-level, independent of the compactification.}

Suppose we have a worldsheet CFT containing a $T^n$ sigma model. We will denote the sigma model fields by $X^\mu$, as is standard in string theory. Let us now orbifold by a finite group $\Gamma$ of isometries of $T^n$ ($\Gamma$ can be abelian or non-abelian). Before orbifolding, the sigma model CFT has a continuous ``momentum'' symmetry which acts by translation (we could tell a completely analogous story for the ``winding'' symmetry). Let $\mathcal{U}_{\delta X}$ denote the invertible topological operator implementing a translation $X^\mu \to X^\mu + \delta X^\mu$. In general, $\mathcal{U}_{\delta X}$ will not be preserved by the $\Gamma$ action, and will be taken to a different translation operator under the action of $g \in \Gamma$. Thus, if we gauge $\Gamma$, the operators $\mathcal{U}_{\delta X}$ will no longer be gauge-invariant.

However, while the operators $\mathcal{U}_{\delta X}$ are not individually gauge invariant, they can still be grouped into orbits $[\delta X]$ of the $\Gamma$ action.\footnote{Be careful: the action of $\Gamma$ on the group of translations is not equal to the action on the sigma model target $T^n$. If $g \in \Gamma$ acts on sigma model fields as $X^\mu \mapsto \Lambda^\mu_\nu X^\nu + X_0^\mu$, then it acts on $\mathcal{U}_{\delta X}$ as $\mathcal{U}_{\delta X} \mapsto g \mathcal{U}_{\delta X} g^{-1} = \mathcal{U}_{\Lambda \cdot \delta X}$. Thus, even if $\Gamma$ acts on the sigma model without fixed points, there may still be fixed points in its action on the symmetry operators $\mathcal{U}_{\delta X}$. For example, if we quotient a square $T^2$ by the $\mathbb{Z}_2$ action $(X, Y) \mapsto (X + \pi, -Y)$ in order to obtain a Klein bottle (as discussed in Appendix \ref{app:break_field}), the action on translation operators is $\mathcal{U}_{(\delta X, \delta Y)} \mapsto \mathcal{U}_{(\delta X, - \delta Y)}$, and the space of orbits of symmetry operators is $S^1 \times (S^1/\mathbb{Z}_2)$, not a Klein bottle.} For each orbit $[\delta X]$, we can define a gauge-invariant topological operator by summing over the orbit
\begin{equation}\label{eq:orbit}
    \mathcal{L}_{[\delta X]} = \bigoplus_{\delta X \in [\delta X]} \mathcal{U}_{\delta X}.
\end{equation}
The quantum dimension $\langle \mathcal{L}_{[\delta X]} \rangle$ is given by the size of the orbit $[\delta X]$.

The collection of operators $\mathcal{L}_{[\delta X]}$ define a non-invertible ``momentum'' symmetry of the toroidal orbifold $T^n/\Gamma$ which is the unbroken piece of the full translation symmetry of the un-orbifolded theory $T^n$. The charged operators include (unnormalized) vertex operators:
\begin{equation}
    \mathcal{O}_{[p]} = \sum_{p \in [p]} e^{i p_\mu X^\mu},
\end{equation}
defined by summing plane waves of definite momentum $p$ over a $\Gamma$-orbit $[p]$ in order to form a $\Gamma$-invariant wavefunction.

What selection rules does the non-invertible symmetry place on correlation functions
\begin{equation}\label{eq:momentum_sphere_correlator}
    \langle \mathcal{O}_{[p_1]} \cdots \mathcal{O}_{[p_n]} \rangle^{T^n/\Gamma}_{\mathbb{S}^2},
\end{equation}
in the orbifolded theory? The answer is simple: the associated selection rules are merely the selection rules coming from conservation of momentum before orbifolding, because the tree-level correlators of untwisted operators are exactly equal to those in the un-orbifolded theory. In more detail, the correlation function \eqref{eq:momentum_sphere_correlator} can only be nonzero if there are representatives $p_i \in [p_i]$ such that
\begin{equation}
    p_1 + \cdots + p_n = 0,
\end{equation}
i.e., such that momentum is conserved. In Appendix \ref{app:selection_rules}, we explain in detail how to re-derive this selection rule using the topological operators \eqref{eq:orbit} for the simple $c = 1$ orbifold $S^1/\mathbb{Z}_2$ (see also \cite{Thorngren:2021yso, Chang:2020imq}).

\begin{figure}[t!]
\begin{center}
\includegraphics[scale = 0.75, trim = {1.5cm 6.0cm 0.0cm 2.0cm}]{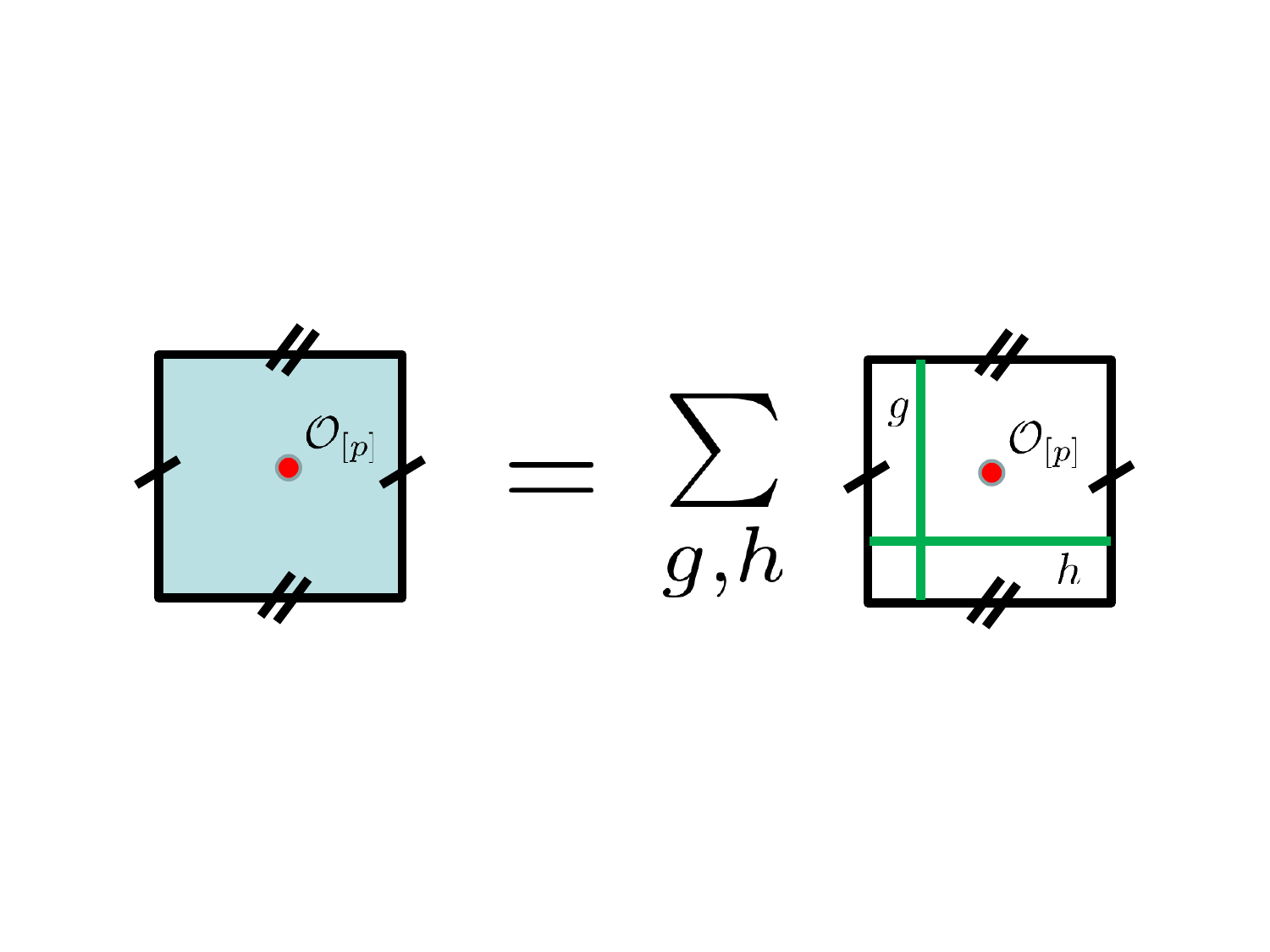}
\caption{The torus one-point function \eqref{eq:momentum_one_point} in the orbifold theory (left, shaded) is computed by summing torus one-point functions in the un-orbifolded theory (right) with insertions of $\Gamma$ symmetry lines (green). The sum runs over all pairs $g, h \in \Gamma$ of commuting elements. Even if the contribution from $g = h = 1$ vanishes, the other terms with nontrivial line insertions may be nonzero. This illustrates that the difference between a theory and its orbifold is simply which topological line insertions are considered to contribute to ``vacuum'' correlation functions.}
\label{fig:torus_one_point_orbifold}
\end{center}
\end{figure}

Suppose we want to calculate the torus correlation function
\begin{equation}
    \langle \mathcal{O}_{[p_1]} \cdots \mathcal{O}_{[p_n]} \rangle_{\mathbb{T}^2}^{T^n/\Gamma}
\end{equation}
in the orbifold theory. For simplicity of the discussion, let us focus on the case of a torus 1-point function
\begin{equation}\label{eq:momentum_one_point}
    \langle \mathcal{O}_{[p]} \rangle_{\mathbb{T}^2}^{T^n/\Gamma}
\end{equation}
Because of the selection rule for the momentum symmetry ``upstairs,'' the torus one-point function $\langle \mathcal{O}_{[p]} \rangle_{\mathbb{T}^2}^{T^n}$ in the un-orbifolded theory must vanish if $p \neq 0$. However, in the orbifold theory, the torus one-point function \eqref{eq:momentum_one_point} involves summing over insertions of commuting pairs of $\Gamma$ symmetry lines on the two cycles of $\mathbb{T}^2$, as illustrated in Figure \ref{fig:torus_one_point_orbifold}. In contrast to the torus one-point function $\langle \mathcal{O}_{[p]} \rangle_{\mathbb{T}^2}^{T^n}$ in the vacuum of the un-orbifolded theory, the torus one-point function of $\mathcal{O}_{[p]}$ in the presence of $\Gamma$ symmetry lines may be nonzero.\footnote{This can be understood by saying that the $\Gamma$ symmetry lines can carry momentum charge, due to their failure to commute with translation.}

\begin{figure}[t!]
\begin{center}
\includegraphics[scale = 0.5, trim = {1.5cm 2.0cm 0.0cm 2.0cm}]{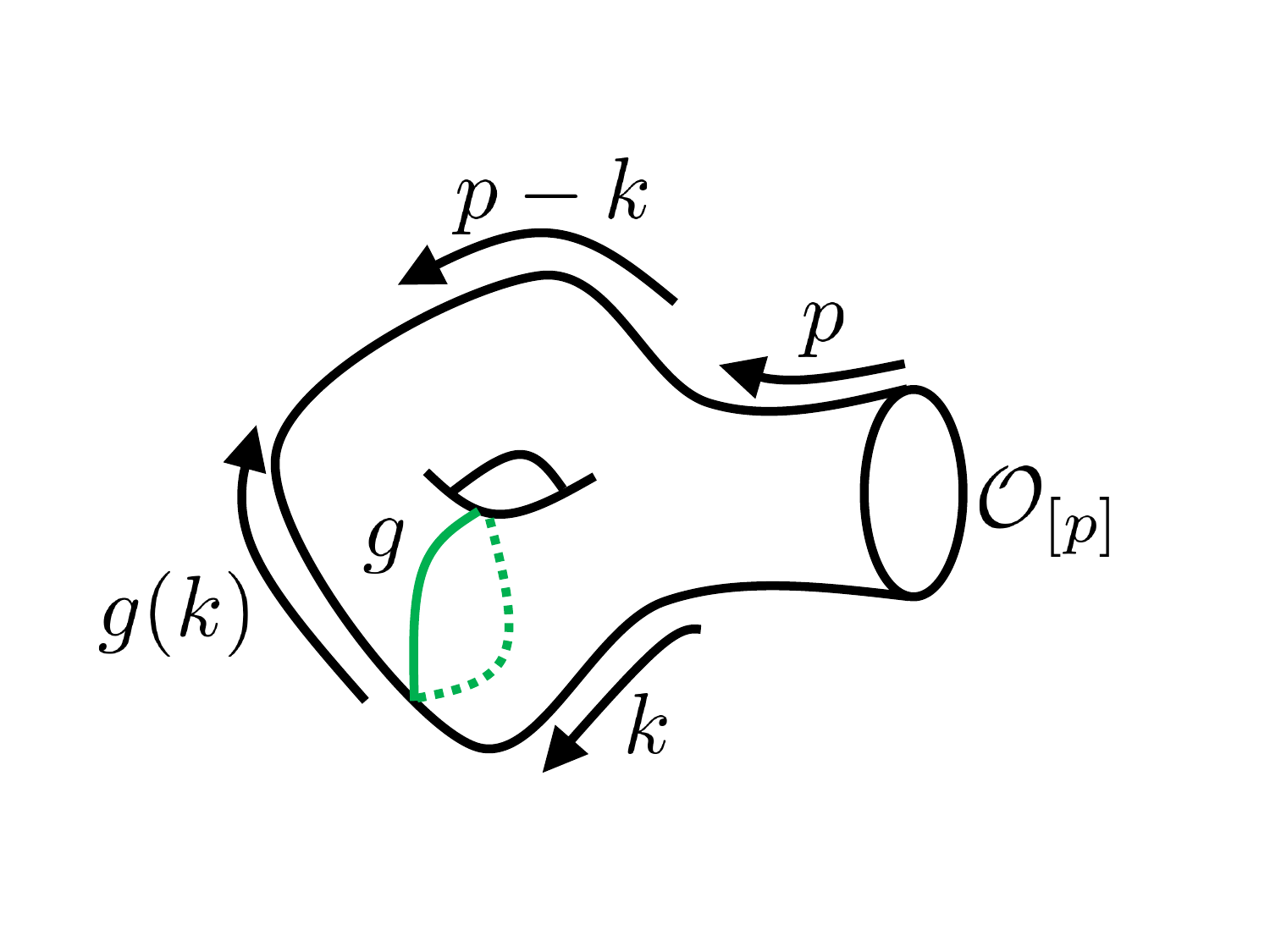}
\caption{Tracking the flow of momentum charge through a torus one-point function of $\mathcal{O}_{[p]}$ with the insertion of a symmetry line $g$ (green). In the orbifold theory, this contributes to the torus one-point function of $\mathcal{O}_{[p]}$ in the orbifold vacuum.}
\label{fig:torusflow}
\end{center}
\end{figure}

To see why this can happen, let us track momentum charge as it flows through the torus with the insertion of a symmetry line for $g \in \Gamma$ (see Figure \ref{fig:torusflow}). Our insertion of $\mathcal{O}_{[p]}$ inserts some momentum $p \in [p]$, which can split into two parts $k, p-k$ running through the two sides of the torus. Before joining, one of the parts, say $k$, is acted on by the $\Gamma$ symmetry line, transforming to some other momentum $g(k)$. Finally, the momentum charge running through the two sides meets, and must annihilate by the selection rules in the un-orbifolded theory. Thus, we have
\begin{equation}\label{eq:momentum_adjoint_subcat}
    p - k + g(k) = 0, \quad \text{or}, \quad p = k + g(-k).
\end{equation}
Thus, any operator $\mathcal{O}_{[p]}$ such that $p = k + g(-k)$ for some $g$ and some $k$ could acquire a nonzero torus one-point function. For example, in the $c = 1$ orbifold $S^1/\mathbb{Z}_2$ by $X \mapsto -X$ discussed in Appendix \ref{app:selection_rules}, the condition \eqref{eq:momentum_adjoint_subcat} is equivalent to the condition that $p$ be even. This remaining selection rule corresponds to an unbroken invertible translation symmetry, given by a $\pi$-rotation of $S^1$.

\begin{figure}[t!]
\begin{center}
\includegraphics[scale = 0.45, trim = {2.0cm 5.0cm 0.0cm 4.0cm}]{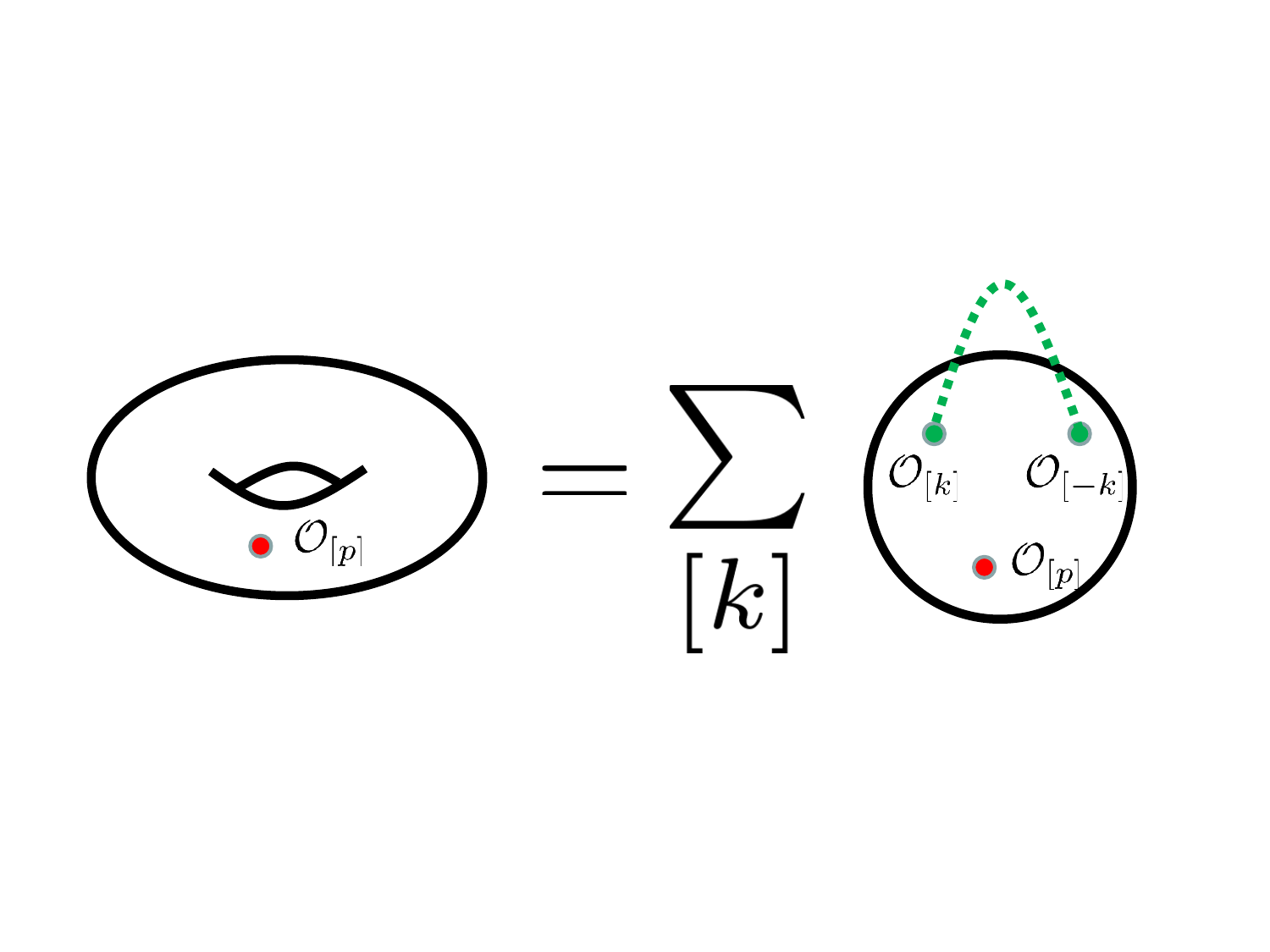}
\caption{A torus one-point function can be written as a sphere three-point function with two points sewn together. The sewing procedure involves a trace over the sewn operators' Hilbert space.}
\label{fig:torus_from_sewing}
\end{center}
\end{figure}

An alternative way to make this argument is to build the torus one-point function \eqref{eq:momentum_one_point} by sewing together sphere three-point functions (see Figure \ref{fig:torus_from_sewing})
\begin{equation}
    \langle \mathcal{O}_{[p]} \mathcal{O}_{[k]} \mathcal{O}_{[-k]} \rangle_{\mathbb{S}^2}^{T^n/\Gamma}.
\end{equation}
Because of the $\Gamma$ orbifold, we have $\mathcal{O}_{[-k]} = \mathcal{O}_{[g(-k)]}$, and so this sphere three-point function can be nonzero if \eqref{eq:momentum_adjoint_subcat} is satisfied. Let us conclude this section by noting that if we define the fusion of orbits $[p_1] \otimes [p_2]$ analogously to the fusion of conjugacy classes (defined in the previous section), then \eqref{eq:momentum_adjoint_subcat} is precisely the condition that $[p]$ appears in a fusion $[k] \otimes [-k]$.

\subsection{General Story}\label{sec:GENERAL}

So far, we have seen that in the example of the $\mathrm{Rep}(\Gamma)$ symmetry of non-abelian orbifolds and in the example of the non-invertible momentum symmetry of toroidal orbifolds, the non-invertible symmetry present at tree level is broken at one loop. In both cases, we saw that the charges $\mu$ of the symmetry that could acquire nonzero one-point functions on the torus $\mathbb{T}^2$ were those that appeared in fusions $\rho \otimes \overline{\rho}$ of some charge with its dual. These charges generate what is known as the ``adjoint subcategory'' of the category of charges \cite[Definition 4.14.5.]{etingof2016tensor} (see also \cite{McNamara:2021cuo}). A natural guess, then, would be that this is the general story.

However, this is certainly wrong, due to the possibility of non-abelian symmetry! For example, rather than a non-abelian orbifold with $\mathrm{Rep}(\Gamma)$ symmetry, consider the un-orbifolded theory itself, with non-abelian $\Gamma$ symmetry. Then, the charges of local operators are given by representations $\mu$ of $\Gamma$. Even if $\mu$ appears in a fusion $\rho \otimes \overline{\rho}$, an operator $\mathcal{O}_\mu$ charged in $\mu$ cannot acquire a nonzero torus one-point function (or, indeed, a one-point function at any order in $g_s$), as it is charged under the invertible symmetry $\Gamma$, whose selection rules hold on any topology.

To see the issue, suppose we try to replicate the argument from the previous section depicted in Figure \ref{fig:torus_from_sewing} for an operator charged under a non-abelian invertible symmetry. Thus, consider a torus one-point function
\begin{equation}\label{eq:non_abelian_one_point}
    \langle \mathcal{O}_\mu^a \rangle_{\mathbb{T}^2},
\end{equation}
of an operator $\mathcal{O}_\mu^a$ charged in a representation $\mu \subset \rho \otimes \overline{\rho}$, where $a = 1, \dots, \mathrm{dim}(\mu)$ is an internal index running over a basis for $\mu$. If we try to build the one-point function \eqref{eq:non_abelian_one_point} by sewing together sphere three-point functions
\begin{equation}\label{eq:non_abelian_three_point}
    \langle \mathcal{O}_\mu^a \mathcal{O}_\rho^b \mathcal{O}_{\overline{\rho}}^{\overline{c}} \rangle_{\mathbb{S}^2},
\end{equation}
we are forced to trace over the internal indices $b, c$ of the charged operators $\mathcal{O}_\rho^b, \mathcal{O}_{\overline{\rho}}^{\overline c}$. While \eqref{eq:non_abelian_three_point} may be nonzero, it will be proportional to the Clebsch–Gordan coefficients $C^{ab\overline{c}}$ for the fusion channel $\rho \otimes \overline{\rho} \to \mu$. But if $\mu$ is a nontrivial irreducible representation, then the trace $\sum_b C^{a b \overline{b}}$ must vanish, so we do not produce a nonzero torus one-point function.

While the ``adjoint subcategory'' of charges appearing in fusions $\rho \otimes \overline{\rho}$ of charges with their duals does not generally describe the subset of charges that can acquire a torus one-point function, it does still have an important meaning: it described the set of charges that can acquire a torus one-point function possibly in the presence of topological line insertions on nontrivial cycles of $\mathbb{T}^2$ (see Appendix \ref{sec:TORUS_ONE_POINT}). This set of charges is invariant under any possible orbifoldings, as the difference between a theory and its orbifold is merely which topological lines we consider ``gauged,'' or ``condensed,'' i.e., part of the vacuum.

However, for the purposes of string theory, where we only sum over worldsheets without the insertion of topological lines, the more refined question of which charges can get a torus one-point function without any line insertions is essential. In Appendix \ref{sec:TORUS_ONE_POINT}, we give a formal characterization of this set of charges. We strongly suspect that this set is always precisely the set of charges needed to break our non-invertible symmetry to its maximal invertible sub-symmetry, but we were not able to give a full proof (outside of special cases such as the Verlinde lines of a diagonal RCFT, see Appendix \ref{sec:DIAGONAL_RCFT}). As evidence for the generality of this claim, we verify it explicitly for the case of a truly exotic non-invertible symmetry (Haagerup) in Appendix \ref{sec:HAAGERUP}.

As a final note, strictly speaking, what we expect in general is that the selection rules of any non-invertible symmetry are not automatically imposed by the symmetry on correlation functions at higher genus. Thus, in the absence of a conspiracy, we expect the symmetry to be broken at $g_s \neq 0$. Of course, one could imagine there might be exceptional string theories where the selection rules of the non-invertible symmetry continue to hold at higher genus due to nontrivial cancellations, even though the they did not have to. In fact, we will see precisely such a case below in section \ref{sec:TENSIONLESS} on the worldsheet of a tensionless (but not infinitely weakly coupled) string theory in AdS.

\section{Non-Invertible Gauge Symmetries in AdS/CFT} \label{sec:ADS}

In the previous section we argued that at least in perturbative string
backgrounds, unbroken non-invertible gauge symmetries coupled to gravity can only arise in a
suitable limit where a tower of light states enter the spectrum.
In this section we explore more general backgrounds in quantum gravity such as
Anti-de Sitter (AdS)\ space using the AdS/CFT\ correspondence.

Indeed, there is a well-studied sense in which non-invertible gauge symmetries can
easily arise in AdS backgrounds. To see why, suppose we have a $D$-dimensional
CFT$_{D}$ with a semiclassical gravity dual. Suppose also that the CFT$_{D}$
has a non-invertible global symmetry. According to the \textquotedblleft
standard rules\textquotedblright\ of the AdS$_{D+1}$/CFT$_{D}$
\ correspondence, any global symmetry of the boundary theory ought to be
gauged in the bulk. From this perspective, we can immediately generate
examples of gauged non-invertible symmetries in the bulk!

To better understand this, it is helpful to briefly review some
aspects of categorical symmetries for a general $D$-dimensional quantum field
theory QFT$_{D}$. One way to capture the categorical symmetries of the QFT$_D$ is
in terms of a $(D+1)$-dimensional topological field theory (TFT) known as the
symmetry TFT (SymTFT).\footnote{For discussion of various aspects of SymTFTs, see e.g.,
references \cite{Reshetikhin:1991tc, Turaev:1992hq, Barrett:1993ab,
Witten:1998wy, Fuchs:2002cm, Kirillov:2010nh, Kapustin:2010if, Kitaev:2011dxc,
Fuchs:2012dt, Freed:2012bs, Freed:2018cec, Apruzzi:2021nmk, Freed:2022qnc,
Kaidi:2022cpf, Baume:2023kkf, Brennan:2024fgj, Heckman:2024oot, Antinucci:2024zjp}.}
Working on a $(D+1)$-dimensional spacetime of the form $\mathbb{I} \times M_{D}$ with $\mathbb{I}$ an interval and $M_{D}$ the original $D$-dimensional spacetime, we specify physical boundary conditions at one end of the interval and gapped (i.e., topological) boundary conditions at the other end. The theory on the
physical boundary conditions is referred to as a \textquotedblleft relative
QFT\textquotedblright\ in the sense of \cite{Freed:2012bs}. The possible global forms of
the QFT$_D$ are specified by the choice of gapped boundary condition.\footnote{There are some subtleties
with imposing such boundary conditions in the case of continuous symmetries, and in the context of holography one ought not impose gapped boundary conditions anyway. For further discussion on this in the context of SymTFTs and holography, see respectively
\cite{Brennan:2024fgj, Antinucci:2024zjp} and \cite{Heckman:2024oot}.}
Contracting the interval then produces an absolute QFT$_D$.
Switching from one choice of boundary conditions to another is
interpreted in the boundary QFT$_{D}$ as the gauging of a non-anomalous (possibly non-invertible) global symmetry.

For a holographic CFT$_{D}$ with a semiclassical AdS$_{D+1}$ dual, one can view the SymTFT$_{D+1}$ as a topological subsector of
the bulk gravitational theory, as has been explicitly verified in a
number of top down constructions \cite{Apruzzi:2022rei,
GarciaEtxebarria:2022vzq, Heckman:2022muc, Heckman:2022xgu, Dierigl:2023jdp,
Cvetic:2023plv, Bah:2023ymy, Apruzzi:2023uma, Cvetic:2023pgm}, as well as from
a bottom up point of view in \cite{Heckman:2024oot}. More precisely, the
SymTFT$_{D+1}$ should be viewed as a small sliver in the bulk AdS$_{D+1}$,
where the physical boundary conditions of the relative QFT\ have now been
\textquotedblleft smeared out\textquotedblright\ over the rest of the
$(D+1)$-dimensional bulk \cite{Heckman:2024oot}. For our present purposes, it is
enough to observe that the bulk AdS$_{D+1}$ has a topological subsector given
by the SymTFT$_{D+1}$.

As an illustrative example, consider 4D $\mathcal{N}=4$ Super Yang-Mills (SYM)
gauge theory with gauge algebra $\mathfrak{su}(N)$. The global form of the gauge group could, in
general, be of the form $SU(N)/\mathbb{Z}_K$ for any $K$ which divides $N$. The 1-form symmetries of this SYM theory are described by the 5D SymTFT\ with topological term of the form:%
\begin{equation}
S_{5D}^{\text{top}}=\frac{N}{2\pi}\int B_{2}\wedge dC_{2},
\end{equation}
where $B_{2}$ and $C_{2}$ are, in the 4D boundary, interpreted as the
background fields for electric and magnetic 1-form symmetries of the theory.
The global form of the SYM gauge group is specified by the boundary conditions for $B, C$: we can specify $B=0$ or $C=0$, or some more general admixture. This topological term
naturally arises in type IIB\ supergravity via the
10D\ topological term $F_{5}\wedge B_{2}\wedge dC_{2}$ reduced over the
$S^{5}$ factor of AdS$_{5}\times S^{5}$ in the presence of $N$ units of $F_5$ flux. As explained in \cite{Aharony:1998qu}, the choice of boundary conditions for this doublet of 2-form potentials
fixes the center of the gauge group on the boundary. More generally, there are
now many known realizations of SymTFTs via string constructions (see e.g.
\cite{Apruzzi:2021nmk, vanBeest:2022fss, Baume:2023kkf}). As a final comment
on this example, observe that gauging the 1-form symmetry of the 4D\ CFT
allows us to switch polarizations, i.e., this corresponds to changing the
gapped boundary conditions of the boundary theory. For example, starting with
$SU(N)$ gauge theory and an electric 1-form symmetry, gauging the 1-form
symmetry produces the $SU(N)/\mathbb{Z}_N$ gauge theory with a magnetic 1-form symmetry.

This simple example describes the SymTFT for an invertible global symmetry of a CFT$_D$ with a holographic dual, but many examples studied in the literature involve the SymTFT of a non-invertible global symmetry of the CFT$_D$ \cite{Apruzzi:2022rei,
GarciaEtxebarria:2022vzq, Heckman:2022muc, Heckman:2022xgu, Dierigl:2023jdp,
Cvetic:2023plv, Bah:2023ymy, Apruzzi:2023uma, Cvetic:2023pgm}. For these cases, one could rightfully describe the bulk SymTFT sector as a non-invertible gauge theory in the conventional sense,\footnote{As noted in the Introduction, this is true in the sense of ``summing over a mesh of topological operators,'' and likely also in the sense of a redundancy of the description assuming the results of \cite{kawagoe2024levinwen} generalize.} and so there are certainly many examples of unbroken non-invertible gauge symmetries in string theory. However, in spite of appearances, in all examples we know how to
explicitly realize, these \textquotedblleft non-invertible
gauge symmetries\textquotedblright\ are of a rather mild type: they could alternatively be described by invertible gauge theories, either in a dual frame or with appropriate Chern-Simons terms.

So far, our discussion has focused on examples of AdS/CFT where the bulk dual is well described by semiclassical Einstein gravity. If we relax this assumption, we can look for examples of non-invertible symmetries in CFTs whose bulk duals are not semiclassical, and which are described by something like a tensionless string theory. We find, in the example of AdS$_{3}\times S^{3}\times T^{4}$ with one unit of NS5 flux \cite{Gaberdiel:2018rqv, Eberhardt:2018ouy, Eberhardt:2019ywk} that the CFT$_2$ admits non-invertible symmetries whose bulk dual is not described by the associated SymTFT. Instead, the bulk dual is the less-benign stringy non-invertible gauge symmetry described in the previous section, realized as non-invertible symmetry on the worldsheet of the tensionless string.

The rest of this section is organized as follows. To illustrate some of the general issues, we first revisit the case of $\mathrm{Rep} (\Gamma)$
symmetries in the special case of the background AdS$_{3}\times S^{3}\times
T^{4}$. In this case, we argue that although the boundary theory admits a
polarization with a global $\mathrm{Rep} (\Gamma)$ symmetry, the bulk theory is
nevertheless captured by a conventional $\Gamma$ gauge symmetry (i.e., an invertible theory),
so in this sense in the bulk we have an invertible symmetry in disguise. We then
turn to the limit captured by a tensionless string, where we see a large $\mathrm{Rep}(S_N)$
non-invertible symmetry, and argue that the bulk dual is a stringy non-invertible gauge symmetry, realized on the worldsheet of the tensionless string. After this, we turn to a broader discussion of AdS$_{D+1}$/CFT$_{D}$ pairs
for $D > 2$, where we typically encounter symmetries whose non-invertibility is of a very mild type. Consolidating these lessons, we put forward some conjectures on non-invertible symmetries motivated by gravity.

\subsection{Example: Semiclassical AdS$_{3}$/CFT$_{2}$}\label{sec:SEMICLASSICAL}

To illustrate some of the general considerations just presented, we now turn
to an explicit example. Consider the Type IIB NS flux background AdS$_{3}\times S^{3}\times
T^{4}$, with its corresponding CFT$_{2}$ dual given by the F1/NS5 system. This configuration can be obtained
from the near horizon limit of coincident $N_{1}$ F1-strings and $N_{5}$
NS5-branes on $\mathbb{R}^{1,1}\times\mathbb{R}^{4}\times T^{4}$, where both
stacks of branes fill the $\mathbb{R}^{1,1}$ factor, and the NS5-branes wrap
the $T^{4}$ factor as well. In the near horizon limit, the string coupling is frozen via the attractor mechanism, and satisfies:
\begin{equation}\label{eq:gs_AdS3}
g_{s}^{2}\sim\frac{N_{5}}{N_{1}}\times\text{Vol}\left(  T^{4}\right)  ,
\end{equation}
to leading order in $1/N_1$.

We briefly note that this is S-dual to a D1/D5 system, although for our
present purposes we will find the F1/NS5 description more convenient. Let us also note that in the
special case of $N_{5}=1$ there is a tensionless worldsheet description of the full 10D
bulk gravity solution, given in \cite{Gaberdiel:2018rqv, Eberhardt:2018ouy, Eberhardt:2019ywk}. This is not a semiclassical gravity theory, but it has the advantage of being a tractable example of an explicit
worldsheet description of the entire bulk. Let us note that instead of
$T^{4}$ we can also consider a K3 surface, and a specific case of interest are limits of K3 realized as orbifolds of $T^{4}$.

We would now like to understand the presence / absence of non-invertible
symmetries in this background, where we work in the large charge / supergravity limit.
To begin, let us determine some of the discrete gauge symmetries in the bulk. Tuning the moduli of $T^{4}$ we can reach special points in
moduli space where the tuned $T_{\text{tuned}}^{4}$ admits a non-abelian
isometry $\Gamma$. So, in addition to the continuous gauge symmetries that arise generically (from translations of $T^{4}$) we see that the 6D\ spacetime AdS$_{3}\times S^{3}$
has a discrete $\Gamma$ gauge symmetry (from discrete isometries of $T_{\text{tuned}}^4$). If we place the usual Dirichlet boundary conditions on AdS$_{3}\times S^{3}\times
T^{4}_{\text{tuned}}$, this leads in the CFT$_{2}$ to a
global, invertible $0$-form $\Gamma$ symmetry. Now, since this 0-form
symmetry is non-anomalous in the 2D\ CFT, it is natural to ask what happens if
we gauge it. This yields another 2D\ CFT which we denote as CFT$_{2}/\Gamma$.
As explained in \cite{Bhardwaj:2017xup, Tachikawa:2017gyf}, and above in section \ref{sec:NONABELIAN}, the theory
CFT$_{2}/\Gamma$ has a 0-form non-invertible symmetry given by $\mathrm{Rep} (\Gamma)$. In
this symmetry category the symmetry operators are labeled by
finite-dimensional representations of $\Gamma$, and there is an accompanying
fusion rule given by tensor products of such representations.

What is the bulk dual description of the CFT$_{2}/\Gamma$? From the
perspective of the accompanying SymTFT$_{3}$, all we have done is modified the
topological boundary conditions for the theory, changing them from Dirichlet to Neumann for the $\Gamma$ gauge fields. Consequently, we conclude that
in the AdS$_{3}$/CFT$_{2}$ pair with 6D\ geometry AdS$_{3}\times S^{3}$ (after
reduction on the $T_{\text{tuned}}^{4}$ factor) we have changed from electric
to magnetic boundary conditions for the bulk $\Gamma$ gauge theory on
AdS$_{3}\times S^{3}$. This example illustrates an important lesson:\ although one may certainly say that \textquotedblleft the $\mathrm{Rep} (\Gamma)$ symmetry is gauged in the
bulk,\textquotedblright\ there is an alternative presentation
of the bulk theory which is an invertible 0-form $\Gamma$ gauge theory.

Do not confuse the bulk dual of CFT$_2/\Gamma$ with the related background
AdS$_{3}\times S^{3}\times\left(  T^4/\Gamma\right)  $. This is related to AdS$_{3}\times S^{3}\times T_{\text{tuned}}^{4}$ example by gauging the
global $\Gamma$ symmetry of the \textit{bulk worldsheet theory} (which has
$\widehat{c}=10$). We emphasize that this gauging operation is not happening
in the boundary CFT$_{2}$ (which has $c\gg1$), nor in the target space quantum gravity theory. In AdS$_{3}\times S^{3}\times\left(  T_{\text{tuned}}^{4}/\Gamma\right)  $, we generically expect the $\mathrm{Rep}(\Gamma)$ symmetry on the worldsheet to be broken by string loops as described in section \ref{sec:2DEXAMPLES}.

Summarizing, we have seen that in an explicit example with a non-invertible
symmetry of the boundary CFT$_{2}$, the bulk description is rather benign: it
is simply a question of how we choose boundary conditions for the bulk theory.

\subsection{Example: Tensionless String in AdS$_3$}\label{sec:TENSIONLESS}

Let us consider more closely the special case $N_{5}=1$ and
$N_{1}=N$, considered extensively in \cite{Gaberdiel:2018rqv, Eberhardt:2018ouy, Eberhardt:2019ywk}. Although there is no semiclassical gravity dual, this
special case admits an explicit bulk worldsheet description as a tensionless string theory and a characterization of the CFT$_{2}$ as the $N$-fold symmetric product orbifold of the $T^{4}$ sigma model, i.e., Sym$^{N}(T^{4})=(T^{4})^{N}/S_{N}$ (where the symmetric group $S_{N}$ acts by permutation). In this case the CFT$_{2}$ admits a $\mathrm{Rep} (S_{N})$ non-invertible symmetry, which emerges precisely at the orbifold point. Because the bulk theory contains a tensionless string, it admits a higher spin gauge symmetry. For details of this Higgsing of
this higher-spin symmetry as we move away from the symmetric orbifold point, see reference
\cite{Gaberdiel:2015uca} (see also \cite{Lerche:2023wkj}).

What is the bulk dual of this enormous $\mathrm{Rep} (S_{N})$ non-invertible symmetry? In \cite{Gaberdiel:2018rqv, Eberhardt:2018ouy}, it was shown that, in the bulk worldsheet theory, the vertex operators in the $w$-th spectrally flowed sectors satisfy the same selection rules as the conjugacy classes of $w$-cycles in $S_N$, in the limit $N \to \infty$. In other words, the bulk worldsheet theory admits a non-invertible $\mathrm{Rep}(S_\infty)$ symmetry,\footnote{We would not expect to see the finiteness of $\mathrm{Rep}(S_N)$ symmetry in the bulk worldsheet theory, since this finiteness is non-perturbative in $1/N$.} described as the $N \to \infty$ limit of $\mathrm{Rep}(S_N)$ symmetry. This non-invertible symmetry on the worldsheet of the bulk string is the holographic dual of the $\mathrm{Rep}(S_N)$ non-invertible global symmetry of the symmetric product orbifold Sym$^{N}(T^{4})$.

This identification raises a puzzle: the bulk string theory, while tensionless, is not infinitely weakly coupled; indeed, $g_s \sim \sqrt{\mathrm{Vol}(T^4)/N}$ to leading order in $1/N$, as in \eqref{eq:gs_AdS3}. So why is it not the case that the $\mathrm{Rep}(S_\infty)$ symmetry broken by string loops, as in our general story in section \ref{sec:2DEXAMPLES}? There, we had noted a possible way to avoid the breaking effect: if the worldsheet CFT correlation functions were subject to some highly nontrivial cancellations. This is exactly what happens on the tensionless string worldsheet: as described in \cite{Eberhardt:2019ywk}, worldsheet correlation functions of the spectrally flowed vertex operators exactly localize on Riemann surfaces that admit a holomorphic branched cover of $\mathbb{S}^2 = \partial \mathrm{AdS}_3$, with branching specified by the charges $w_i$ of the operator insertions. This localization exactly imposes the selection rules of $\mathrm{Rep}(S_\infty)$ non-invertible symmetry at any order in the string loop expansion, since $\pi_1(\mathbb{S}^2) = 0$ and the only interchange of sheets comes from operator insertions.\footnote{This argument will fail if we consider a more general hyperbolic 3-manifold than AdS$_3$, for example a handlebody of higher genus. However, we do not expect the selection rules to hold even in the non-gravitational CFT$_2$ when we place it on a boundary manifold of nontrivial topology.} For a concrete example, no vertex operator with $w > 1$ can acquire a one-point function at any genus, since there are no holomorphic branched covers of $\mathbb{S}^2$ with only one branch point besides the identity map, which can be viewed as having ``trivial branching'' $w = 1$ (see e.g. \cite[Equation 8.14]{Eberhardt:2019ywk}).

\subsection{Weak Invertibility}\label{sec:WEAK}

Recently a number of examples of non-invertible symmetries in holographic CFT$_D$ for $D>2$ have been discussed, along with their string theory realization in the bulk \cite{Apruzzi:2022rei, GarciaEtxebarria:2022vzq, Heckman:2022muc, Heckman:2022xgu, Dierigl:2023jdp, Bah:2023ymy, Apruzzi:2023uma, Cvetic:2023pgm, Antinucci:2022vyk}.\footnote{See e.g., \cite{Cordova:2022ruw, Schafer-Nameki:2023jdn, Bhardwaj:2023kri, Luo:2023ive, Brennan:2023mmt, Shao:2023gho} for reviews discussing non-invertible symmetries in $D > 2$ in general.} As discussed above, this means that there is a conventional non-invertible gauge symmetry in the bulk gravity dual, described by the SymTFT for the non-invertible symmetry.

However, an important caveat is that all of these examples are
\textquotedblleft weakly invertible,\textquotedblright\ in the sense that
the non-invertibility in their fusion rule only includes defects supported on lower-dimensional subspaces, i.e., a condensate (see e.g. \cite{Gaiotto:2019xmp, Roumpedakis:2022aik}). More explicitly, suppose we have a non-invertible symmetry defect $\mathcal{N}$ supported on a $q$-dimensional subspace. We say $\mathcal{N}$ is \textit{weakly invertible},\footnote{Note that a ``weakly invertible'' symmetry need not be invertible, following the convention that the ``weak'' version of a property does not imply the unmodified version.} or \textit{invertible up to condensates}, if the fusion of $\mathcal{N}$ with $\mathcal{N}^\dagger$ on any given $q$-manifold $N$ satisfies
\begin{equation}\label{eq:weak_invertible}
\mathcal{N}(N)\otimes\mathcal{N}(N)^\dagger=\underset{\mathcal{M}, M}{\sum} \mathcal{M}(M),
\end{equation}
where the sum runs over some topological operators $\mathcal{M}$ supposed on submanifolds $M \subset N$ of strictly lower dimensions $\mathrm{dim}(M) < q$. If we have two such operators $\mathcal{N}_i, \mathcal{N}_j$, their fusion can be described as
\begin{equation}
\mathcal{N}_{i}(N)\otimes\mathcal{N}_{j}(N)=\mathcal{N}_{ij}(N)\otimes\underset{\mathcal{M}_{ij}, M}{\sum} \mathcal{M}_{ij}(M),
\end{equation}
where $\mathcal{N}_{ij}$ is another weakly invertible operator of the same dimension, and $\mathcal{M}_{ij}$ runs over some set of topological operators of lower dimension.\footnote{To prove this, it suffices to show that $\mathcal{N}_i \otimes \mathcal{N}_j$ is irreducible. Suppose it were not, so we had $\mathcal{N}_i \otimes \mathcal{N}_j = \mathcal{A} \oplus \mathcal{B}$. Then by fusing with $\mathcal{N}_j^\dagger$, we would have $\mathcal{N}_i \otimes (\text{condensates}) = (\mathcal{A} \otimes \mathcal{N}_j^\dagger) \oplus (\mathcal{B} \otimes \mathcal{N}_j^\dagger)$. Since reducibility is invariant under fusing with condensates, we would learn that $\mathcal{N}_i$ were reducible, which is incompatible with the weak invertibility of $\mathcal{N}_i$. By the same argument, the fusion of any irreducible operator with a weakly invertible operator is irreducible.} Note that if we ignore condensates, the fusion of weakly invertible operators defines a group law.

Weakly invertible operators are to be contrasted with e.g., the case of Verlinde lines of a 2D RCFT,
where the fusion products involve multiple summands of topological defect lines of the same dimension.
One of the general lessons from top down realizations of non-invertible symmetries is that bulk dual of weakly invertible symmetries is the well-known process of brane / anti-brane annihilation \cite{Apruzzi:2022rei, Bah:2023ymy}, which produce lower-dimensional branes that were dissolved in the original brane / anti-brane pair via the dielectric-brane effect \cite{Myers:1999ps}.\footnote{For explicit examples in AdS/CFT, see e.g., references \cite{Apruzzi:2022rei, GarciaEtxebarria:2022vzq, Heckman:2022muc, Antinucci:2022vyk, Heckman:2022xgu, Dierigl:2023jdp, Bah:2023ymy, Heckman:2024oot}.} The corresponding bulk gauge theory is an invertible gauge theory with triple Chern-Simons terms turned on (see also \cite{Heidenreich:2020pkc}), which capture the possibility of branes of different dimensions to dissolve into one another. Again, we see that the bulk dual of boundary non-invertible symmetry can be rephrased in terms of an invertible gauge theory, now with nontrivial topological couplings. Notably, in this case, we do not have to perform any electro-magnetic duality in the bulk: the bulk gauge fields are invertible, but their electric symmetries are rendered non-invertible by the triple Chern-Simons terms \cite{Choi:2023pdp, Reece:2023iqn, Cordova:2023her, Heidenreich:2023pbi}.\footnote{\label{footnote:long_range} This illustrates a general pattern: strictly speaking, the bulk dual of a global symmetry in the CFT is the approximate electric global symmetry of the bulk gauge fields (see \cite{Heckman:2024oot} for a recent discussion of the approximate bulk symmetry operators arising from the boundary symmetries). This was referred to as ``long range gauge symmetry'' in \cite{Harlow:2018tng}; see also \cite{Cordova:2022rer} for the approximate non-invertible electric symmetry of a non-abelian gauge theory in Maxwell phase.}

There may be more general notions of weak-invertibility beyond the definition \eqref{eq:weak_invertible}, whose bulk duals correspond to invertible symmetries directly, without switching to a magnetic duality frame. For example, suppose we had a CFT$_D$ with a global $O(2) = U(1) \rtimes \mathbb{Z}_2^\mathrm{C}$ symmetry. If we gauge $\mathbb{Z}_2^{\mathrm{C}}$, we obtain a continuous non-invertible symmetry, with topological operators $\mathcal{L}_\theta$ defined as in Appendix \ref{app:selection_rules} (see also \cite{Chang:2020imq, Thorngren:2021yso, Heidenreich:2021xpr, Bhardwaj:2022yxj} for more discussion of this construction). In the bulk, where we have a dynamical $O(2)$ gauge theory, this corresponds to switching our boundary conditions for the $\mathbb{Z}_2^\mathrm{C}$ discrete gauge field, while leaving the Dirichlet boundary conditions for $U(1)$ unchanged. Thus, while we have switched to a magnetic duality frame for $\mathbb{Z}_2^\mathrm{C}$, the $U(1)$ gauge field is directly holographically dual to the non-invertible symmetry in the CFT$_D$, without any electromagnetic duality.\footnote{As in Footnote \ref{footnote:long_range}, what is really happening is that the approximate electric 1-form symmetry of $O(2)$ gauge theory in the bulk is non-invertible \cite{Heidenreich:2021xpr}.}

From the perspective of the fusion algebra of symmetry operators in the CFT$_D$, what is happening is that we have a continuous family of topological operators $\mathcal{L}_\theta$ such that the limit $\theta \to 0$ of $\mathcal{L}_\theta$ is a condensate.\footnote{More precisely, it is the condensate of the dual $(D-2)$-form symmetry generated by topological Wilson lines for $\mathbb{Z}_2^\mathrm{C}$ \cite{Chang:2020imq, Thorngren:2021yso, Heidenreich:2021xpr, Bhardwaj:2022yxj}.} In particular, this means that these operators can be written as
\begin{equation}\label{eq:non_inv_noether}
    \mathcal{L}_\theta(N) = \mathrm{Condensate}(N) \otimes \exp\left( i \theta \int_N \star J\right),
\end{equation}
where $J$ is a 1-form conserved current operator that is only well-defined along the condensate (see also \cite{Thorngren:2021yso, Choi:2022jqy, Cordova:2022ieu, Damia:2022bcd}). The formula \eqref{eq:non_inv_noether} defines a more general sort of ``weak invertibility'' or ``invertibility up to condensates'' beyond \eqref{eq:weak_invertible}, whose bulk dual involves a photon field which is itself charged under some other discrete gauge symmetry (in this case, $\mathbb{Z}_2^\mathrm{C}$). It would be very interesting to determine the most general notion of ``weak invertibility'' in QFT (see section \ref{sec:FUTURE} for further comments on this question).

\subsection{Conjectures Motivated by Gravity}

The general lesson from these examples is that while we do expect a gravity
dual for non-invertible symmetries in theories with a semiclassical bulk, the bulk description is typically
\textquotedblleft benign\textquotedblright, and can be rephrased as a more
conventional invertible gauge theory description, either in an electromagnetically dual frame, or directly as-is in the case of weak invertibility. Motivated by these considerations, it is natural to ask whether the SymTFT$_{D+1}$
for any QFT$_D$ (whether or not it has a semiclassical gravity dual) can always
be presented as a more conventional invertible gauge theory with appropriate topological couplings (i.e., Chern-Simons terms). This is a broader QFT question, but the evidence we have from holographic examples suggests that this more general statement might be true.

Gravity also suggests that there may end up being an upper bound on the number of separate operators that can appear in the fusion products
\begin{equation}
\mathcal{N}_{i}\otimes\mathcal{N}_{j}=\underset{k=1}{\overset{k_{\mathrm{max}%
}}{\sum}}\mathcal{T}_{ij}^{k}\mathcal{N}_{k}.
\end{equation}
of topological operators in holographic CFTs with semiclassical bulk duals. It is tempting to conjecture that $k_{\mathrm{max}} \sim O(1)$ for non-invertible symmetries which are dual to conventional non-invertible gauge symmetries in actual UV complete backgrounds.\footnote{This statement could possibly be extended to the general case by placing a bound on the bulk EFT cutoff.} Indeed, returning to the examples presented in section \ref{sec:SEMICLASSICAL}, the collection of non-abelian orbifolds $\Gamma$ which can serve as isometries of an explicit tuned $T^4$ is rather small, and the resulting dimensions of irreducible representations is also quite limited. Indeed, the only case we saw with a possibly large number of fusion products involved $S_N$, whose bulk dual was realized as a stringy non-invertible gauge symmetry in a tensionless string theory. Observe that in other AdS backgrounds such as AdS$\times S^n / \Gamma$ (see \cite{Collins:2022nux}), the order of $\Gamma$ can be arbitrarily large, but the dimensions of the irreducible representations (and thus the number $k_\mathrm{max}$ of fusion products) are far smaller. Clearly, it would be interesting to see whether there exist holographic CFTs having non-invertible symmetries with large $k_{\mathrm{max}}$, or conversely, whether there is a nontrivial Swampland constraint.

\section{Approximate Non-Invertible Symmetries in the String Landscape} \label{sec:SWAMP}

Despite being broken in spacetime, in this section we will show that non-invertible symmetries on the worldsheet can still have interesting implications for effective fields theories arising from string theory. First, we will explain the interplay of non-invertible symmetries with several Swampland constrains, including the Distance conjecture and the Sub(Lattice) Weak Gravity Conjecture.

\subsection{Existence of Towers of States}

In section \ref{sec:continuous_noninvertible} we showed the generic presence of non-invertible symmetries at $g_s=0$ whenever we have string theory compactified on toroidal orbifolds such as $S^1/\mathbb{Z}_2$. More generically, there are non-invertible symmetries present in any orbifold $\mathcal{M}/G$ where $\mathcal{M}$ is a smooth manifold with isometries broken by the $G$ action. This includes many examples with fixed points (including toroidal orbifold Calabi-Yau manifolds); in these cases, there is a question of whether the non-invertible symmetries may be explicitly broken by turning on deformations corresponding to marginal twisted sector operators localized at the fixed points. All of the above arguments, however, also apply to the case when $G$ is freely acting. In these cases, the non-invertible symmetry is exact classically for any choice of geometric moduli, and as discussed in the previous subsection, is only broken by quantum effects at $g_s \neq 0$. Examples of these manifolds include Riemmann and Ricci-flat examples, such as freely acting quotients of $T^n$ and quotients of the form
\begin{equation} \frac{K3\times T^k}{\mathbb{Z}_n},\end{equation}
where the $\mathbb{Z}_n$ is a common subgroup of isometries. Some of these constitute examples of Calabi-Yau manifolds with infinite fundamental group \cite{Hashimoto:2014oma,Hashimoto:2015zqm}.

As we have seen, these non-invertible symmetries are only approximate unless we are in a decompactification or perturbative string limit. A natural question is then whether it plays any role in the physics close, but not exactly at the asymptotic/perturbative limits of effective field theory. We will now show that this non-invertible symmetry can be used to prove the existence of a tower of states that becomes light at infinite field distance, as predicted by the Distance Conjecture \cite{Ooguri:2006in}. Using non-invertible symmetries we can therefore extend the range of asymptotic limits where a proof of the Distance Conjecture is available.\footnote{Of course another way to argue that in the large radius limit for arbitrary compactifications the light KK tower is related to gauge symmetry, as is anticipated by Weak Gravity Conjecture \cite{Arkanihamed:2006dz}, is to note that in this limit we get approximate translational symmetries which lead to gauge symmetries broken by $1/R$ effects.} Importantly, for the first time, the argument does not use the existence of an unbroken gauge symmetry, as is the case in the usual perturbative string \cite{Lee:2018urn} and complex structure moduli space examples \cite{Grimm:2018ohb,Gendler:2020dfp}.  Since the non-invertible symmetry is not exact, we can expect small corrections to the mass and lifetime of the particles in the tower. However the existence of the tower itself in the asymptotic limit is guaranteed by the non-invertible symmetry.

The argument we have in mind is a minor modification of the proof in \cite{Heidenreich:2016aqi, Montero:2016tif} of the Sublattice Weak Gravity Conjecture, which itself is a direct application of spectral flow.\footnote{See also \cite{Heidenreich:2024dmr} for extremely recent progress in this direction.} We will now briefly review the argument in \cite{Heidenreich:2016aqi,Montero:2016tif} (which was itself described inline in \cite{Arkanihamed:2006dz}), in the particular case of a single $U(1)$ gauge field and then explain how it gets modified for the case of a non-invertible symmetry.

Consider a 2D worldsheet CFT with an invertible $U(1)$ symmetry generated by a holomorphic current $j$ at level $N$. In other words,
\begin{equation} j(z) j(0)\sim \frac{N}{z^2}\end{equation}
In these circumstances, one may consider the partition function with complex chemical potential
\begin{equation} Z(\mu,\tau) \equiv \text{Tr} ( q^{L_0} \bar{q}^{\bar{L}_0} e^{2\pi i Q \mu}),\quad q\equiv e^{2\pi i\tau}. \label{pf0} \end{equation}
This partition function transforms covariantly under $SL(2,\mathbb{Z})$
\begin{equation}Z\left(\frac{\mu}{c\tau+d},\frac{a\tau+b}{c\tau+d}\right)=e^{i\pi N\frac{c\mu^2}{c\tau+d}}\, Z(\mu,\tau)\label{sl2d}  \end{equation}
(the lack of exact invariance is due to the anomalous conservation of the holomorphic current, see e.g. \cite{Kraus:2006wn}) and it also satisfies
\begin{equation} Z(\mu,\tau)=Z(\mu+1,\tau)\label{e433}\end{equation}
since the charges are quantized. As shown in \cite{Heidenreich:2016aqi,Montero:2016tif}, imposing \eqref{e433} and \eqref{sl2d} together implies that the whole spectrum of CFT operators is invariant under a simultaneous shift
\begin{equation} h\,\rightarrow h+\frac{Q^2}{2N},\quad Q\,\rightarrow\, Q+N,\label{spefl}\end{equation}
known as a spectral flow automorphism. Equation \eqref{spefl} implies that the CFT spectrum arranges itself into towers of states where the $U(1)$ charge shifts by $N$ and the dimensions increase accordingly. In particular, the tower associated to the identity is labeled by a parameter $k$ and has charges $Q=kN$ and weight $h=Nk^2/2$.
In a perturbative string context, after level-matching, these operators correspond to a tower of particles with mass and $U(1)$ charge
\begin{equation}\label{mQ} m\sim \sqrt{N} k,\quad Q=kN\end{equation}
which exactly saturates or satisfies the Sublattice Weak Gravity Conjecture \cite{Heidenreich:2016aqi,Montero:2016tif}. More to the point, this tower of states becomes light in the perturbative string limit: the proof of the sublattice WGC is also a particular case of the Distance Conjecture. Although we reviewed here the case of a single holomorphic current, the setup is general, applying to any number of abelian currents of any chirality.

The basic point of this subsection is that the above argument goes through almost unchanged in the case where the symmetry is non-invertible. Imagine gauging a $\mathbb{Z}_2$ symmetry that sends $j$ to $-j$, as would be the case when going from $S^1$ to the $S^1/\mathbb{Z}_2$ sigma model of section \ref{sec:continuous_noninvertible}. The partition function \eqref{pf0} is no longer a well-defined object, but the quantity
\begin{equation} \tilde{Z}(\mu,\tau)\equiv Z(\mu,\tau) +Z(-\mu,\tau)   \end{equation}
is,\footnote{In the orbifold theory, it is the partition function in the untwisted sector with an insertion of a line for the quantum symmetry with complex potential. We are allowed to ignore the contribution of the twisted sector in the orbifold theory since it is charged under the quantum $\mathbb{Z}_2$ symmetry and we can consider only the uncharged states.} and clearly, it inherits the modular transformation properties \eqref{sl2d}, leading to the existence of a non-invertible version of spectral flow (notice that the weight $h$ in \eqref{spefl} is invariant under sign flip of $Q$). Therefore, we obtain again a tower of particles in the spacetime, which are just the particles from the tower in the unorbifolded tower which were not projected out. Importantly, in the orbifold case they are not charged under any massless gauge field.

In the particular case of an $S^1/\mathbb{Z}_2$ sigma model, what the spectral flow automorphism predicts is precisely the tower of interval KK modes. More generally, for $\mathcal{M}/G$, the tower of states thus predicted is that of KK modes. Although the existence of these states was known directly from a bulk EFT analysis, it is interesting to see it arise purely from a non-invertible symmetry. The fact that one has a worldsheet argument implies that similar towers can be obtained for winding modes and non-geometric models, too.  Moreover, it supports the idea of interpreting the tower of states as a quantum gravity obstruction to restoring a global symmetry at infinite distance \cite{Grimm:2018ohb,Gendler:2020dfp,Cordova:2022rer}. We have seen that these non-invertible symmetries are restored at weak string coupling. A preliminary argument in Appendix \ref{app:break_field} suggests that they also seem to be restored at large radius, even without having a string worldsheet description, and would become exact at infinite field distance in these decompactification limits. In all known examples in string theory, the towers of the Distance Conjecture are either KK modes or string modes \cite{Lee:2019wij}, getting light and weakly coupled asymptotically.  For the latter case, it seems we can sometimes identify a weakly broken symmetry that becomes exact as the string coupling vanishes. For the former case, any translational diffeomorphism of the higher-dimensional vacuum corresponds to a sort of approximate symmetry from the lower-dimensional perspective that gets restored upon decompactification. However, in most cases, these symmetries are already broken at classical level, unlike the non-invertible symmetry that is only broken by loop effects. If the compact manifold has some isometry, this yields a continuous gauge symmetry in the lower dimensional EFT (which would become global at infinite distance unless there is a KK tower of states signaling decompactification of extra dimensions). In the absence of an isometry, we can still have in certain cases an approximate non-invertible symmetry that is preserved at classical level and only broken by quantum corrections. It would be interesting to see whether this weakly broken non-invertible symmetry can be generalized to other decompactification limits beyond toroidal orbifolds.

\subsection{Interplay with (Sub)Lattice WGC}

The non-invertible symmetry also has implications for the sublattice WGC described in the previous subsection. As shown in equation \eqref{mQ}, the states shown to exist via spectral flow only have charge given by a multiple of $N$, the level of the $U(1)$ current algebra. The charged states therefore only live in a sublattice. Since the value of $N$ is unconstrained, this sublattice can be made arbitrarily sparse, in principle, and the Swampland implications get correspondingly diluted; interesting Swampland statements are about constraining the spectrum of light states, while for large enough $N$, the states predicted by spectral flow can become arbitrarily massive. This undesirable feature of the sublattice version of WGC is known as the ``loophole'' in the literature \cite{Rudelius:2015xta,Montero:2015ofa,Brown:2015lia,Harlow:2022ich}.\footnote{This loophole and a related construction\cite{Wen:1985qj} was already noted in \cite{Arkanihamed:2006dz} and was the basis of the observation in the original paper that the WGC does not always hold for the minimally charged state in the theory.}

Explicit examples realizing this loophole are known \cite{Heidenreich:2016aqi}, though they all take the form of freely acting orbifolds of $T^n$, just like the ones discussed above. Although in the covering $T^n$ there are $n$ currents realized at level 1, after orbifolding it is possible to obtain currents that are realized at higher level, resulting in a sublattice of states.  In those cases, though, we have seen that there is still a non-invertible global symmetry that survives at tree level in the worldsheet. We will now explain how the interplay between this non-invertible and invertible currents can be used to improve on the sublattice WGC, to conclude that even for non-superextremal states there are light charged states (though they are not superextremal).

\begin{figure}[ht!]
    \centering
    \includegraphics[scale=0.25]{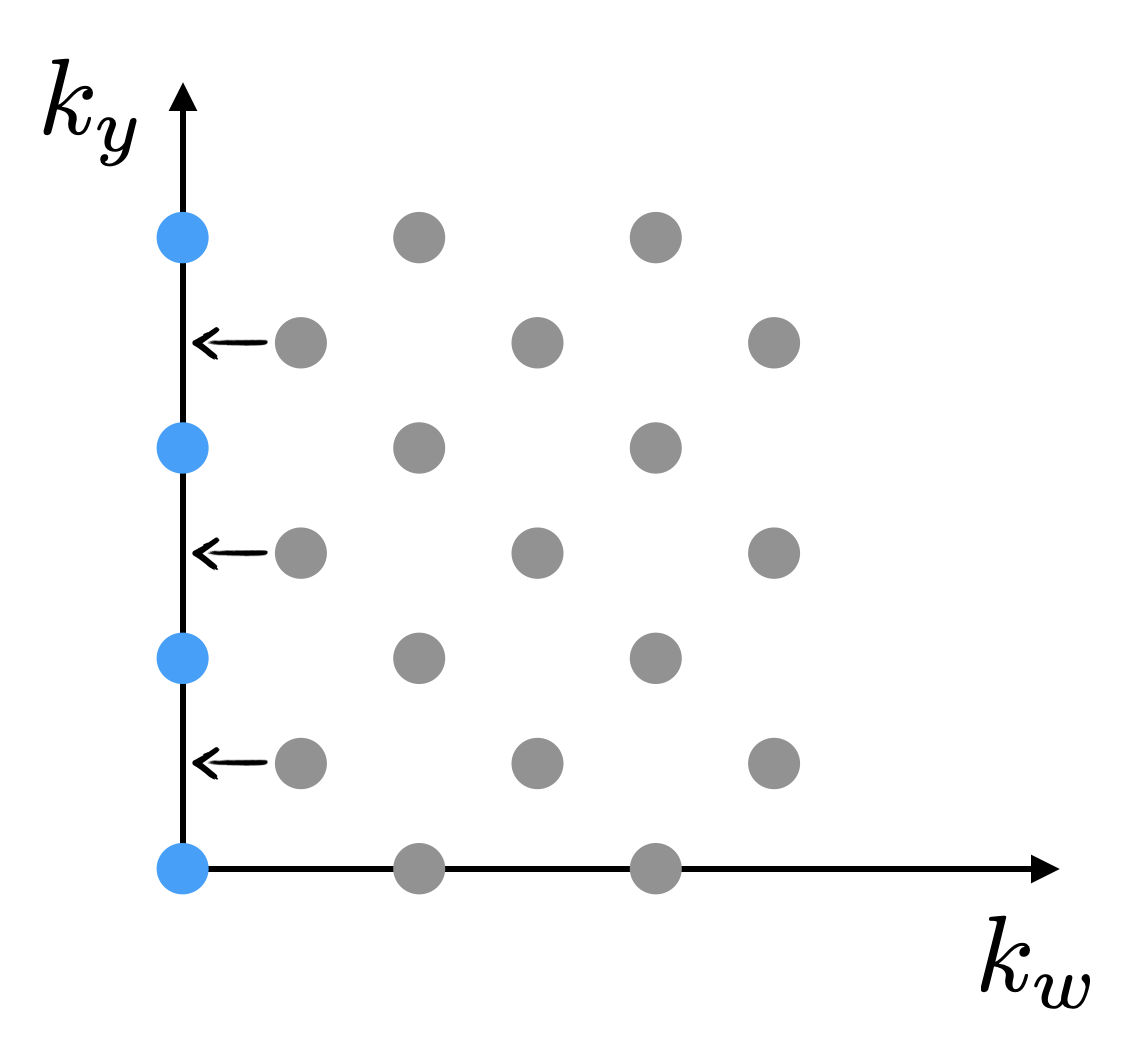}
    \caption{Two dimensional slice of the lattice of KK momenta. Only the vertical axis is associated to a U(1) gauge charge. The blue dots are extremal states (saturating the WGC) while the grey ones are subextremal.}
    \label{fig:lat}
\end{figure}

To illustrate this, let us consider for example the orbifold $T^3/\mathbb{Z}_2\times \mathbb{Z}_2'$ discussed in \cite{Heidenreich:2016aqi}, where the freely acting group acts as
\begin{align}
\mathbb{Z}_2\,:\ \theta_w\rightarrow \theta_w+\pi,\,\theta_y\rightarrow \theta_y+\pi,\\
\mathbb{Z}'_2\,:\ \theta_w\rightarrow -\theta_w,\,\theta_z\rightarrow \theta_z+\pi.
\end{align}
The unorbifolded toroidal compactification contains three massless gauge fields, but the  $\mathbb{Z}'_2$ projects out the first one associated to the direction $w$. Hence, the charge lattice is only given by KK charges $(k_y,k_z)$. In what follows we will focus in the subspace with $k_z=0$ for simplicity, but the lessons are general. Even in this subspace, we have KK modes with non-vanishing momentum $k_w$; Notice that the first $\mathbb{Z}_2$ implies that unprojected KK modes with odd values of $k_y$ must also have an odd value of $k_w$. The corresponding charge lattice is represented in Figure \ref{fig:lat} as dots in a two dimensional slice. The KK masses and charges are given by
\begin{equation}
m^2=\frac{k_w^2}{R_w^2}+\frac{k_y^2}{R_y^2}+\frac{k_z^2}{R_z^2}\ , \quad gQ=(k_y,k_z)\frac1{R}
\label{psoe}\end{equation}
where we have also included the KK gauge coupling $1/R$.

 As emphasized above, only the vertical direction in the figure corresponds to the charge direction of the massless gauge field that survives the orbifold action, so only $k_y$ (and not $k_w$) is truly a gauge charge.  Hence, we need to project all KK states over the vertical direction to obtain the set of charged states under the massless gauge field. The KK states with $k_w=0$ and even $k_y$ are extremal (i.e. saturate the WGC bound) since $|Q|=M$ (there are lattice sites on the vertical axis). However, states with odd $k_y$ and odd $k_w$ are subextremal (i.e. violate the WGC bound) since  $|Q|<M$ due to the $\frac{k_w^2}{R_w^2}$ contribution on the mass. Therefore, there is only a sublattice of states satisfying the WGC, as noticed in \cite{Heidenreich:2016aqi}.

Our main point here is that the modular flow argument for non-invertible symmetries of the previous section allows us to recover the existence of the full lattice of KK towers, including those with non-vanishing $k_w$ that are not charged under any massless gauge field, even though they will be charged under some discrete gauge symmetries. Non-invertible spectral flow predicts the existence of states with masses and charges given by \eqref{psoe} for all allowed values of $k_w,k_y,k_z$.  Moreover, it allows us to quantify how much these subextremal particles are violating the WGC. In this particular case we have that
 \begin{equation}
 \frac{m^2}{Q^2}=1+\frac{k_w^2}{k_y^2} \frac{R_y^2}{R_w^2}>1.
 \end{equation}
We see that the violation of the Lattice WGC becomes negligible for very large charges $k_y$; while it depends on the ratio of the radii for small charges. Hence, even if the first light state satisfying the WGC does not have the minimal possible value of the charge (the WGC states start with charge 2 in this example), we can still use spectral flow of the non-invertible symmetry to show the existence of light states of unit charge whose charge to mass ratio is constrained and which only mildly violate the WGC. For small charges, the charge-to-mass ratio is of order one, so in any case, the  dangers of the sublattice WGC loophole for phenomenological applications in this example is significantly ameliorated; we still get a full lattice of light (in terms of the gauge coupling) charged states, even if they are not exactly superextremal.

However, part of the reason why one still gets light states is that in this example, the sublattice is of index 2 (half the lattice sites contain superextremal particles). It is possible to consider bottom-up constructions \cite{Heidenreich:2016aqi} where the index of the sublattice becomes very large; in these examples, the first values of the charge may still contain very heavy states.\footnote{Indeed, there are open string examples where the index of the sublattice can be made very large, provided that a sufficiently long warped throat exists in a Calabi-Yau with fluxes.} It therefore remains an essential question to bound the index of the lattice in general. A related question suggested by the above analysis is whether all examples with an abelian current at level $N$ contain additional non-invertible symmetries that allow one to predict the masses of subextremal states.

\subsection{Supersymmetric Protection}\label{sec:SUSY}

The paper \cite{Palti:2020qlc} discusses an interesting phenomenon in certain SUGRA theories, where observables which are generically nonvanishing were shown to vanish exactly, in violation of the naturalness principle. This violation could be avoided in the presence of higher SUSY, but this higher SUSY was not observed in these examples. This observation led to the ``supersymmetric genericity conjecture,'' which states that such protections must be a result of the theory being related to a higher-SUSY theory in some indirect way. In practice, in all of the examples considered, the lower-SUSY theory was related to the higher-SUSY theory through the gauging of a discrete subgroup of the R-symmetry. The examples in the rest of this section follow from the discussions in \cite{Palti:2020qlc}.

Our discussion of non-invertible symmetries gives a new perspective on this observation. The construction in section \ref{sec:continuous_noninvertible} applies here as well, since certain supercharges appear to be projected out when we gauge a discrete subgroup of the R-symmetry, but instead they reappear as non-invertible symmetries.\footnote{On the worldsheet, the construction is analogous to \eqref{eq:orbit} for toroidal orbifolds: one can take sums of the worldsheet symmetry operators for target space SUSY over an orbit of the orbifold group. This definition is likely easier in the Green-Schwarz formalism, where target space SUSY is manifest on the worldsheet.} Naturalness is thus not violated if we generalize its definition to include non-invertible symmetries.

For concreteness we discuss a specific example. In 4D $\mathcal{N}=2$ SUGRA, the prepotential $\mathcal{F}$ for a vector multiplet contains a polynomial term in the superfields (of degree at most 3) and exponential terms generated by worldsheet instanton effects,
\begin{equation}
    \mathcal{F}=\mathcal{F}_{\text{polynomial}}\left(\Phi\right)+\sum_{n,i}B_{n}\left(\Phi\right)e^{-a_{n}^{i}\Phi_{i}}\;.
\end{equation}
The coefficients $B_n$ are generically nonzero, although if SUSY is enhanced to $\mathcal{N}=4$ the coefficients $B_n$ vanish for all $n$. We can now discuss an explicit example of an apparent violation of naturalness. Consider Type II string theory on orbifolds of $\mathbb{T}^6$ or $K3\times\mathbb{T}^2$. If the orbifold preserves $\mathcal{N}\geq4$ SUSY, then the coefficients $B_n$ vanish automatically. However, these coefficients vanish also if the orbifold preserves only $\mathcal{N}=2$ SUSY. The reason for this is that the prepotential is computed using only the genus-zero contribution on the worldsheet. As a result, amplitudes of the orbifold theory restricted to untwisted operators are identical to that of the unorbifolded theory, and so restricting $\mathcal{F}$ to the untwisted fields gives only a cubic term. If the twisted sectors do not include massless fields, the prepotential will then be exactly cubic.

As discussed above, this violation of naturalness is avoided since the supercharges which are projected out under orbifolding still leave behind a non-invertible symmetry and its corresponding sphere selection rules. While these are broken at leading order in $g_{s}$, they are preserved at genus 0, and as a result they constrain tree-level amplitudes of untwisted operators just like their invertible versions in the unorbifolded theory.\footnote{It is crucial that $\mathcal{F}$ is computed at genus 0, where the non-invertible symmetry is unbroken. In an analogous computation for the heterotic string on $\mathbb{T}^6/\Gamma$, the prepotential also receives one-loop corrections, and as a result the non-invertible SUSY is not enough to protect it, and indeed one finds non-vanishing $B_n$'s.} Then the puzzling protection of $\mathcal{F}$ in this example can be attributed to non-invertible symmetries.

\section{Discussion and Future Directions} \label{sec:FUTURE}

In this paper we have studied potential realizations of gauged non-invertible symmetries in quantum gravity. In cases where we have a global non-invertible symmetry in the worldsheet CFT, we found that higher-loop effects generically break the putative non-invertible symmetry to its maximal invertible sub-symmetry, unless there are nontrivial cancellations as in the case of the tensionless string considered in section \ref{sec:TENSIONLESS}. As a result, the target space physics of non-invertible worldsheet symmetry, which we referred to as ``stringy non-invertible gauge symmetry,'' is broken (Higgsed) away from the tensionless limit. Moreover, in CFTs with a semiclassical holographic dual, we found that the bulk dual of non-invertible symmetry was always ``benign,'' in the sense that it could alternatively be characterized by an invertible gauge theory in one way or another, while in CFTs whose dual contained a tensionless string, we recovered an unbroken stringy non-invertible gauge symmetry.

How should we think about stringy non-invertible gauge symmetry? It is something beyond any gauge symmetry that can be described in effective field theory. One answer would be that it should be viewed as a non-invertible extension of the gauge symmetries of string field theory, which are generically Higgsed when the string acquires a tension. This would give a natural explanation for the fact that stringy non-invertible gauge symmetries are restored only in the limit of a tensionless string.

To support this picture, consider the continuous non-invertible symmetry in the simple orbifold $S^1/\mathbb{Z}^2$, studied in detail in Appendix \ref{app:selection_rules}. What is the corresponding gauge boson? On the worldsheet, the current for the invertible $U(1)$ symmetry of the $S^1$ sigma model has been projected out, and only survives as a disorder operator attached to a topological line. Thus, it cannot be used to create an on-shell state in the string theory spectrum. However, perhaps such disorder operators could be included in string field theory as a way to describe the Higgsed gauge bosons for stringy non-invertible gauge symmetry. A caveat is that these disorder operators appear for any value of $g_s$, so such an interpretation would also require understanding their role at nonzero $g_s$ where the non-invertible symmetry is broken.

As a further comment, consider again the mesh of $\mathcal{N} \otimes \mathcal{N}^\dagger$ that appears on a Riemann surface when we sweep a non-invertible operator $\mathcal{N}$ across, as discussed in section \ref{sec:SELECTIONRULES}. This mesh has a very natural interpretation: the fusion $\mathcal{N} \otimes \mathcal{N}^\dagger$ has the structure of a Frobenius algebra, meaning it corresponds to a (possibly non-invertible) orbifolding of the worldsheet theory, defined exactly by inserting it as a mesh (see e.g., \cite{Bhardwaj:2017xup}). Moreover, $\mathcal{N} \otimes \mathcal{N}^\dagger$ is manifestly Morita trivial, meaning that the worldsheet theory we obtain by orbifolding is the original worldsheet theory! Thus, in general, a non-invertible operator tells us that the string theory at hand is a self-orbifold, possibly under a non-invertible orbifolding \cite{TY:1998, Thorngren:2021yso, Choi:2023vgk}.

As a result, one way to interpret the results of this paper is that ``the orbifold procedure'' itself should be viewed as part of the gauge symmetries of string field theory. In a sense, it is then very odd that non-invertible symmetries are generically broken at $g_s \neq 0$, since the orbifold procedure (even by non-invertible symmetries) still makes sense order-by-order in string perturbation theory. One way to resolve the tension is to note that the orbifold procedure is acting on the string background: a fixed 2D CFT, viewed as a solution to the classical string theory equations of motion with $g_s = 0$. Thus, it makes sense to study string perturbations about this background to any loop order, even if those same loop effects lead to a breaking of the corresponding non-invertible symmetry at $g_s \neq 0$.

The low-energy limit of string theory is the effective field theory of supergravity. Therefore, whenever the relevant string states survive the field theory limit, it might be possible to reinterpret the one-loop effects that break non-invertible worldsheet symmetries as a sum of ordinary field theory diagrams. For example, if we consider a toroidal orbifold as in section \ref{sec:continuous_noninvertible} with no fixed points, in the large-volume limit, the effects from twisted sectors running in the loop will be suppressed, so the breaking can be understood entirely from Kaluza-Klein modes running in the loop. In this sense, it may be possible to see non-invertible symmetries emerging not just in limits with a tensionless string, but also in decompactification limits, where the lower dimensional EFT breaks down to be replaced by a higher-dimensional one.
We give some preliminary comments in this direction in Appendix \ref{app:break_field}.

In this paper, we described how a stringy realization of non-invertible symmetries arises near the perturbative string limit where the symmetry gets approximately restored.  It would be interesting to better understand the fate of these symmetries in other tuned backgrounds. For example, one might naively expect that M-theory compactified on a non-abelian orbifold would have a non-invertible 1-form symmetry arising from the non-invertible 1-form symmetry on the M2-brane worldvolume theory. However, this symmetry seems to be badly broken,\footnote{This fits with the fact that there is no small parameter controlling any sum over M2-brane worldvolume topologies, which could have suppressed the symmetry breaking effect from topologically nontrivial configurations.} unless we put the theory on a further circle, where in the limit of small radius it would lead to an approximate non-invertible 0-form symmetry as realized on the dual Type IIA string theory.  It would be interesting to find further evidence for this expectation or to find evidence to the contrary.

As a practical comment, let us note that the breaking of non-invertible worldsheet symmetries away from the tensionless limit does not mean they are useless: since they are still good approximate symmetries, they can be used to constrain the spectrum and interactions of the theory. We have shown how non-invertible symmetries are able to fill in the gaps left by the usual worldsheet derivation of the sublattice WGC, and more generally, how they can be used to predict the existence of towers of states which are not charged under any continuous gauge symmetry, which is of interest for the Distance conjecture. Moreover, the existence of certain examples which exhibit properties as if they had higher supersymmetry can be attributed to the presence of a non-invertible fermionic symmetry (i.e., a $\mathbb{Z}_{2}$ odd internal symmetry) on the worldsheet. The examples we considered here can also be understood via more elementary techniques; the role of non-invertible symmetries here is to provide a new perspective on an old physical phenomenon.

While our discussions have been phrased in the context of quantum gravity, there may be general lessons for non-invertible symmetries in $D$-dimensional QFTs without gravity. In particular, in section \ref{sec:ADS}, we saw that the only example of a non-invertible symmetry which could not be viewed as ``weakly invertible'' in one sense or another\footnote{Meaning that the bulk dual is directly an invertible gauge theory without switching duality frames, as discussed in section \ref{sec:WEAK}. The general QFT definition of this property is still unclear.} was the case of global categorical symmetries of a CFT$_2$. In this case, the bulk dual description in AdS$_3$ could be described as a non-abelian gauge theory after performing electromagnetic duality. Notably, the magnetically charged objects under a discrete non-abelian gauge symmetry have codimension 2. Now, in any dimension $D$, we can always find $(D-2)$-form global symmetries that cannot be viewed as ``weakly invertible'' simply by considering $\Gamma$ gauge theory for a non-abelian group $\Gamma$. However, it is possible that in any QFT$_D$, the only non-invertible symmetries which cannot be viewed as ``weakly invertible'' must be $p$-form symmetries for $p \geq D - 2$.\footnote{In fact, one could view \cite{Johnson-Freyd:2020ivj} as establishing something like this in $D = 3$.} This statement must be restricted to local QFT, as we have discussed examples of 0-form stringy non-invertible gauge symmetries in dimensions $D > 3$ which cannot be viewed as ``weakly invertible,'' such as the symmetries restored as $g_s \to 0$ in 6D from non-abelian orbifolds $T^4/\Gamma$.

Lastly, let us mention that in the context of holography with a semiclassical bulk gravity, we did not find \textit{any} examples of non-invertible symmetries that could not be regarded as invertible gauge symmetries in the bulk with appropriate choices of boundary conditions and topological terms. It would be interesting to study whether this pattern holds up in general, and whether there are nontrivial constraints on the non-invertible symmetries of holographic CFTs coming from constraints on UV complete quantum gravity.

\section*{Acknowledgements}

We thank B. Rayhaun for collaborating in early stages of this project. We thank
L. Bhardwaj, C.-M. Chang, C. C\'ordova, A. Debray, D. Delmastro, L. Eberhardt, B. Haghighat, M. H\"ubner, C. Murdia, J. Parra-Martinez, K. Roumpedakis, S. Seifnashri, R. Thorngren, X. Yu, and Y. Zheng for helpful
discussions. The work of JJH is supported by DOE (HEP) Award DE-SC0013528, a University
Research Foundation grant at the University of Pennsylvania and by BSF grant 2022100.
JM is supported by the U.S. Department of Energy, Office of Science, Office of High Energy Physics, under Award Number DESC0011632.
MM is supported by an Atraccion del Talento Fellowship 2022-T1/TIC-23956 from Comunidad de Madrid. The work of IV is also partly supported by the grant RYC2019-028512-I from the MCI (Spain) and the ERC Starting Grant QGuide-101042568 - StG 2021. MM and IV also acknowledge  the support of the grants CEX2020-001007-S and PID2021-123017NB-I00, funded by MCIN/AEI/10.13039/501100011033 and by ERDF A way of making Europe. CV is supported by a grant from the Simons Foundation (602883,CV), the DellaPietra Foundation, and by NSF grant PHY-2013858. We thank the 2023 Simons Summer Workshop for hospitality. JM, MM, AS and IV thank ICBS Daejeon for hospitality during the ``CERN-Korea workshop: Recent trends in and out of the Swampland''. JM, MM and IV thank the Aspen Center for Physics, which is supported by National Science Foundation grant PHY-2210452 for hospitality during the 2023 summer program ``Traversing the Particle Physics Peaks — Phenomenology to Formal.'' JJH and JM thank the 2023 NYU Satellite Workshop on Global Categorical Symmetries for hospitality during part of this work. JJH, JM, and AS thank the 2023 meeting of the Simons Collaboration on Global Categorical Symmetries for hospitality during part of this work.

\appendix



\section{Selection Rules for $S^1/\mathbb{Z}_2$}\label{app:selection_rules}

In this Appendix we illustrate the discussion in section \ref{sec:continuous_noninvertible} in the simple torodial orbifold $S^1/\mathbb{Z}_2$, and explicitly re-derive the selection rules on sphere correlation functions from the non-invertible symmetries.

\subsection{Non-Invertible Symmetries of $S^1/\mathbb{Z}_2$}

We start with the unorbifolded theory $S^1$ with coordinate $X \sim X + 2 \pi R$, which is the $c=1$ compact boson CFT with radius $R$. We will discuss only the momentum symmetry, but there is a completely analogous discussion for the winding symmetry as well. We have local momentum vertex operators
\begin{equation}
\mathcal{O}_{m}(z, \overline{z})=e^{i m X(z, \overline{z})/R}.
\end{equation}
The $U(1)$ momentum symmetry is generated by the invertible TDLs
\begin{equation}
\mathcal{U}_{\theta} = \exp \left[\frac{i \theta R}{2 \pi} \int \star d X\right],
\end{equation}
whose action on local operators is
\begin{equation}
    \mathcal{U}_{\theta}:  \mathcal{O}_{m} \mapsto e^{im \theta}\mathcal{O}_{m}\;.
\end{equation}

Now, we consider the orbifold by $X\to -X$, under which we have
\begin{equation}
    \mathcal{O}_m \to \mathcal{O}_{-m}, \quad \mathcal{U}_\theta \to \mathcal{U}_{-\theta}.
\end{equation}
In the orbifolded theory, the spectrum of local operators consists of twisted and untwisted sectors. In the untwisted sectors we have vertex operators
\begin{equation}
    \mathcal{O}^+_{m}=\frac{1}{\sqrt 2}(\mathcal{O}_{m}+\mathcal{O}_{-m})\;,
\end{equation}
We also have a topological, invertible Wilson line for the gauged $\mathbb{Z}_2$ symmetry, which we will denote by $\eta$, and which implements the quantum $\mathbb{Z}_2$ symmetry. At the end of $\eta$, we have a sector of disorder operators, which includes the gauge-non-invariant operators
\begin{equation}
    \mathcal{O}^-_{m}=\frac{i}{\sqrt 2}(\mathcal{O}_{m}-\mathcal{O}_{-m})
\end{equation}
which have been projected out of the spectrum of local operators.\footnote{We have included a factor of $i$ so that $\mathcal{O}^-_m$ descends from a Hermitian operator in the un-orbifolded theory.}

In the orbifolded theory, the $U(1)$ momentum operators are generically projected out, leaving only the identity operator $\mathcal{U}_0$ and the half-shift operator $\mathcal{U}_\pi$, which corresponds to the invertible $\mathbb{Z}_2$ symmetry that flips the interval $S^1/\mathbb{Z}_2$. However, using the construction in equation \eqref{eq:orbit}, we have a continuum of non-invertible TDLs, labeled by $\theta \in (0, \pi)$ (see \cite{Thorngren:2021yso, Chang:2020imq, Heidenreich:2021xpr})
:
\begin{equation}
    \mathcal{L}_{\theta}= \mathcal{U}_{\theta} \oplus \mathcal{U}_{-\theta}\;.
\end{equation}
The action of these non-invertible lines on the physical degrees of freedom actually has a simple intuitive description: we first ``un-orbifold,'' then rotate $S^1$ by an angle $\theta$ (or $- \theta$, either will work), then ``re-orbifold,'' obtaining $S^1/\mathbb{Z}_2$ again, but folded along an axis rotated by $\theta$.\footnote{This intuitive language can be made precise by noting that $\mathcal{L}_\theta = \mathcal{D} \otimes \mathcal{U}_\theta \otimes \mathcal{D}^\dagger$, where $\mathcal{D}$ is the topological gauging interface from $S^1$ to $S^1/\mathbb{Z}_2$, i.e., the Dirichlet boundary condition for the $\mathbb{Z}_2$ gauge fields.}

\begin{figure}[t!]
     \centering
     \begin{subfigure}[b]{0.6\textwidth}
         \centering
         \includegraphics[width=\textwidth]{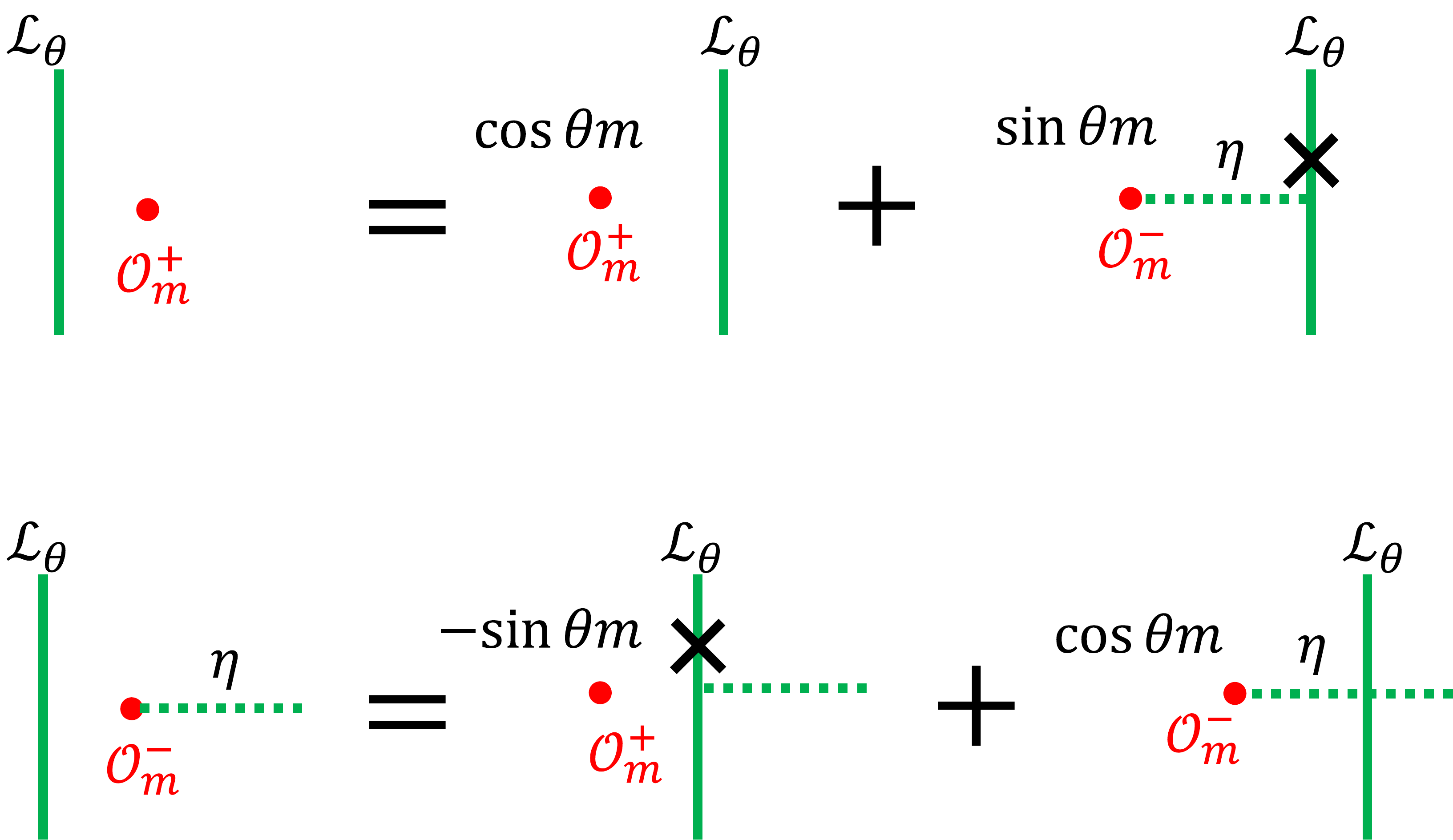}
         \caption{ }
            \end{subfigure}
     \hfill
     \begin{subfigure}[b]{0.7\textwidth}
         \centering
         \includegraphics[width=\textwidth]{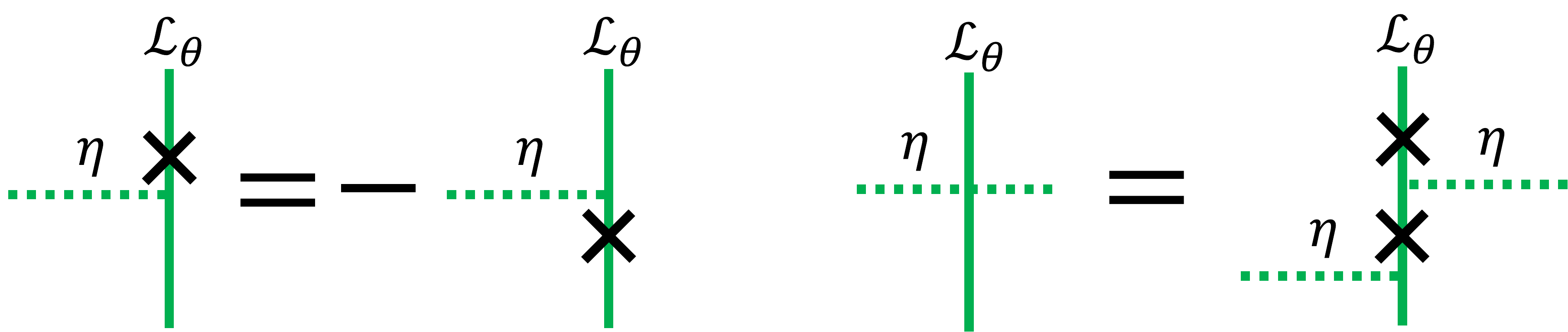}
         \caption{ }
     \end{subfigure}
     \hfill
        \caption{(a) Rules for sweeping a TDL past vertex operators. A black ``X'' is used to denote the orientation of a junction (see e.g. \cite{Chang:2018iay}). (b) Conventions for the orientation of the junctions.}
        \label{fig:sweeping_vertex_ops}
\end{figure}

As described above, the action of non-invertible symmetries by sweeping generally maps local operators (or disorder operators) to linear combinations of local operators and disorder operators (see Figure \ref{fig:sweeping_vertex_ops}). We have
\begin{equation}\label{eq:cos_action}
\begin{split}
    \mathcal{L}_{\theta} : \;&\mathcal{O}^+_{m} \mapsto \cos(m \theta)\ \mathcal{O}^+_{m} + \sin(m \theta)\ \mathcal{O}^-_{m} \;,\\
    \mathcal{L}_{\theta} : \;&\mathcal{O}^-_{m} \mapsto - \sin(m \theta)\ \mathcal{O}^+_{m} + \cos(m \theta)\ \mathcal{O}^-_{m} \;.
    \end{split}
\end{equation}
Their fusion rules are given by
\begin{equation}\label{eq:fusion}
    \mathcal{L}_{\theta}\mathcal{L}_{\theta'}=\mathcal{L}_{\theta+\theta'}+\mathcal{L}_{\theta-\theta'}\;,
\end{equation}
where we have defined the reducible lines
\begin{equation}
    \begin{split}
    \mathcal{L}_{0}=1+\eta\;,\qquad & \mathcal{L}_{\pi}=(1+\eta)\otimes\mathcal{U}_\pi\;,
    \end{split}
\end{equation}
to simplify notation.

\subsection{Selection Rules}

We can now discuss selection rules on the sphere. Following section \ref{sec:SELECTIONRULES}, a naive attempt would be to consider some correlation function
\begin{equation}
    \langle \mathcal{O}^+_{m_1}(x_1)...\mathcal{O}^+_{m_n}(x_n)\rangle\;,
\end{equation}
and then generate a selection rule by nucleating a non-invertible line $\mathcal{L}_\theta$ and sweeping it past the operators (see figure \ref{fig:selection_rules}). This fails, since the action of the non-invertible lines in \eqref{eq:cos_action} inevitably maps this correlator to correlators involving $\mathcal{O}^-_m$. We are thus instead forced to consider all correlators of the form
\begin{equation}
    \langle \mathcal{O}^\pm_{m_1}(x_1)...\mathcal{O}^\pm_{m_n}(x_n)\rangle\;,
\end{equation}
and see how the non-invertible lines relate them.

For simplicity we start with the general two-point functions $\langle \mathcal{O}^+_{m}\mathcal{O}^+_{m'} \rangle\;,\langle \mathcal{O}^-_{m}\mathcal{O}^-_{m'}  \rangle$, where we suppress the position dependence. Following the procedure in figure \ref{fig:selection_rules}, one finds the following set of equations:
\begin{equation}
\left(
\begin{array}
[c]{rr}%
1-\cos(\theta m)\cos(\theta m') & -\sin(\theta m)\sin(\theta m')\\
-\sin(\theta m)\sin(\theta m') & 1-\cos(\theta m)\cos(\theta m')
\end{array}
\right)
\left(
\begin{array}
[c]{r}%
\langle \mathcal{O}^+_m \mathcal{O}^+_{m'}\rangle\\
\langle\mathcal{O}^-_m \mathcal{O}^-_{m'}\rangle
\end{array}
\right) =0\;.
\end{equation}
These equations have a nontrivial solution for generic $\theta$ only if the determinant of the matrix vanishes. Its eigenvalues are
\begin{equation}
    E_{\pm}=1-\cos(\theta m)\cos(\theta m')\pm \sin(\theta m)\sin(\theta m')=1-\cos(\theta(m \pm  m'))\;.
\end{equation}
The nonzero correlators are thus only those with $m=\pm m'$, as expected.

Next we consider three-point functions. A similar analysis results in the set of equations
\begin{equation}
    (1+M)
    \left(\begin{array}[c]{r}
        \langle \mathcal{O}^+_{m}\mathcal{O}^+_{m'}\mathcal{O}^+_{m''}\rangle \\
        \langle \mathcal{O}^+_{m}\mathcal{O}^-_{m'}\mathcal{O}^-_{m''}\rangle \\
        \langle \mathcal{O}^-_{m}\mathcal{O}^+_{m'}\mathcal{O}^-_{m''}\rangle \\
        \langle \mathcal{O}^-_{m}\mathcal{O}^-_{m'}\mathcal{O}^+_{m''}\rangle
    \end{array}\right)=0\;,
\end{equation}
where $M$ is the matrix (using $c\equiv \cos,\; s\equiv \sin$)
\begin{equation}
    \footnotesize{\left(\begin{array}[c]{rrrr}
        -8c(\theta m)c(\theta m')c(\theta m'')&-2c(\theta m)s(\theta m')s(\theta m'')&-2c(\theta m')s(\theta m)s(\theta m'')&-2c(\theta m'')s(\theta m')s(\theta m) \\
        -2c(\theta m)s(\theta m')s(\theta m'')&-8c(\theta m)c(\theta m')c(\theta m'')&2c(\theta m'')s(\theta m')s(\theta m)&2c(\theta m')s(\theta m)s(\theta m'') \\
        -2c(\theta m')s(\theta m)s(\theta m'')&2c(\theta m'')s(\theta m')s(\theta m)&-8c(\theta m)c(\theta m')c(\theta m'')&2c(\theta m)s(\theta m')s(\theta m'') \\
        -2c(\theta m'')s(\theta m')s(\theta m)&2c(\theta m')s(\theta m)s(\theta m'')&2c(\theta m)s(\theta m')s(\theta m'')&-8c(\theta m)c(\theta m')c(\theta m'')
    \end{array}\right)}
\normalsize
   \;.
\end{equation}
The eigenvalues of $1+M$ are
\begin{equation}
    E_{s_1,s_2}=1-\cos(\theta (m+s_1 m'+s_2m''))\;,\qquad  s_1,s_2\in\{\pm 1\}.
\end{equation}
 so we find that a nontrivial correlator must have $m+s_1 m'+s_2m''=0$ for some choice of $s_1,s_2$, again as expected. The general result for longer correlators follows from repeated application of this method.

Finally, we explain why this procedure fails to generate selection rules on higher-genus surfaces. For simplicity we put the theory on the torus $\mathbb{T}^2$. The main caveat in the process appears in the last step of Figure \ref{fig:selection_rules}, where we annihilate the topological line ``at infinity''. On $\mathbb{T}^2$, this means taking the fusion of the two lines which meet from opposite ends of each cycle. For invertible lines, this fusion gives the identity and so annihilates the lines, leading to a selection rule. However, for non-invertible lines the result is more complicated. Using the fusion rule \eqref{eq:fusion} we find
\begin{equation}
    \mathcal{L}_\theta \mathcal{L}_{-\theta}=1+\eta+\mathcal{L}_{2\theta}\;.
\end{equation}
Importantly, in addition to the contribution from the identity, non-invertibility forces other contributions to appear. As a result, the process for generating selection rules on the sphere fails to generate a selection rule on the torus (and similarly for higher-genus surfaces). Instead, this process relates a correlation function in the vacuum to the same correlation function but with the topological line $1+\eta+\mathcal{L}_{2\theta}$ wrapping every 1-cycle of the manifold. We thus cannot extract a selection rule from this procedure.

\section{General Torus One-Point Functions}\label{sec:TORUS_ONE_POINT}

In this appendix, we give a general characterization of the set of charges described in section \ref{sec:GENERAL} that can acquire a torus one-point function, both with and without the insertion of topological lines wrapping nontrivial cycles. Recall from \ref{sec:SELECTIONRULES} that if we have a 2D CFT with fusion category $\mathcal{C}$, the set of charges of local and disorder operators are described by representations of the tube algebra $\mathrm{Tube}(\mathcal{C})$, or equivalently, by objects in the Drinfeld center $\mathcal{Z}(\mathcal{C})$ (for a physicist-friendly discussion, see \cite{Lin:2022dhv}). For our discussion here, it will be helpful to recall the 3D (SymTFT) perspective on symmetries of 2D CFTs (again, see \cite{Lin:2022dhv}). To any 2D CFT with symmetry $\mathcal{C}$, we can define a boundary condition $\mathcal{B}$ of the 3D Turaev-Viro (Levin-Wen) TQFT \cite{TV:1992, Levin:2004mi} associated to $\mathcal{C}$, which we will denote $\mathrm{TV}_{\mathcal{C}}$.\footnote{$\mathrm{TV}_\mathcal{C}$ can be viewed as ``$\mathrm{Tube}(\mathcal{C})$ gauge theory,'' due to very recent work \cite{kawagoe2024levinwen}.} We can recover our original 2D CFT by dimensionally reducing $\mathrm{TV}_{\mathcal{C}}$ on an interval, with the physical boundary condition $\mathcal{B}$ on one end, and the topological Dirichlet boundary condition for $\mathcal{C}$ on the other. The category of bulk anyons is given by the Drinfeld center $\mathcal{Z}(\mathcal{C})$. We can form local and disorder operators in our 2D CFT by stretching anyon line operators across the interval, possibly attached to a topological line running along the Dirichlet boundary.

\begin{figure}[t!]
\begin{center}
\includegraphics[scale = 0.5, trim = {1.5cm 3.0cm 0.0cm 2.0cm}]{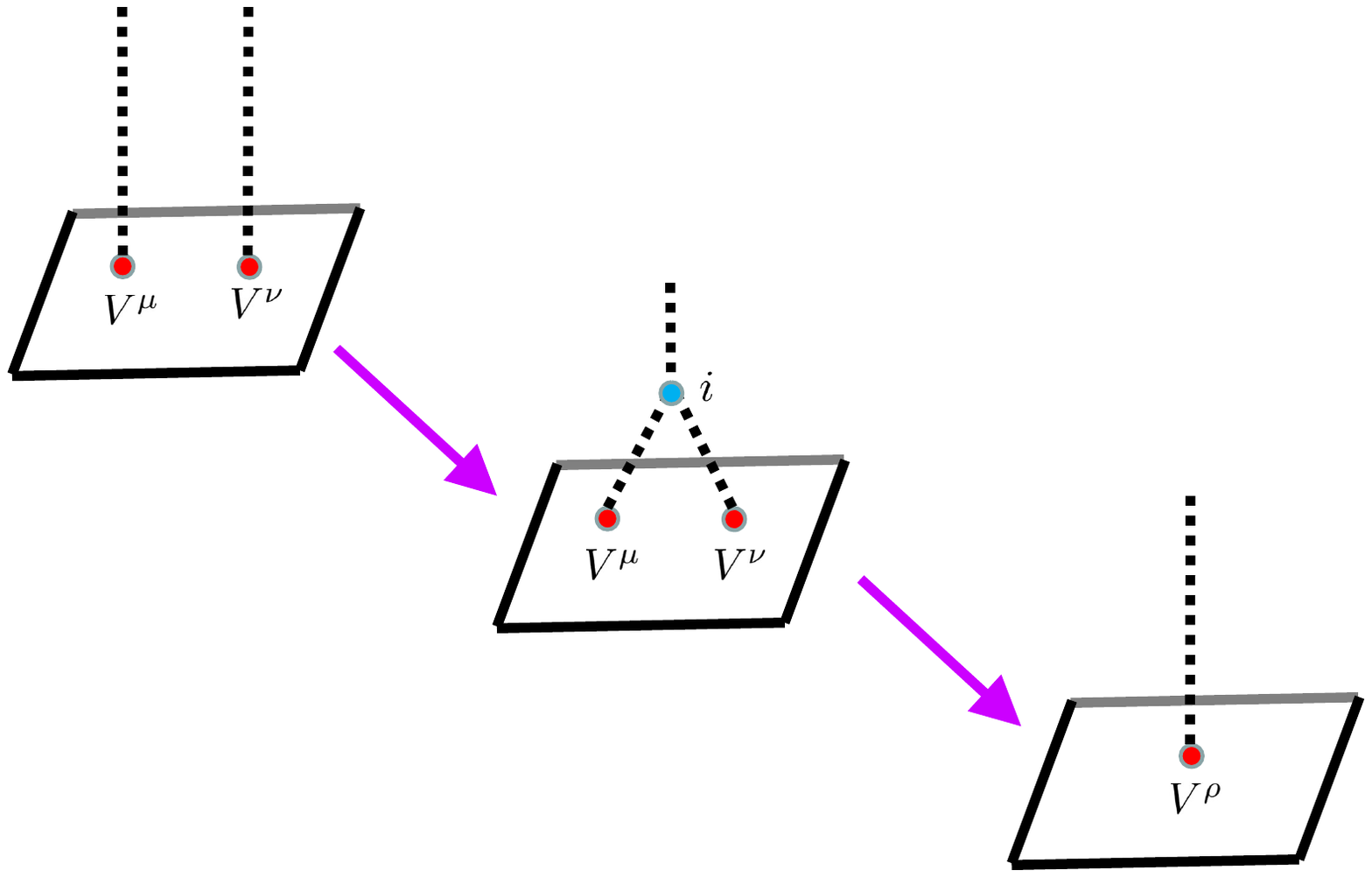}
\caption{Starting with two topological junctions in $V^\mu, V^\nu$ and a fusion channel $i : \mu \otimes \nu \to \rho$, we can describe the algebra multiplication maps $\left(m_{\mu \nu}^{\rho}\right)_i$ as follows. First, we fuse the anyon lines with the bulk junction operator corresponding to our chosen fusion channel. We then shrink the three junction operators (two boundary, one bulk) to a point, obtaining some topological junction in $V^\rho$.}
\label{fig:condensable_algebra_structure}
\end{center}
\end{figure}

The Dirichlet boundary condition is associated to a Lagrangian algebra $\mathcal{A} \in \mathcal{Z}(\mathcal{C})$, which describes the anyons that are condensed on the boundary (can end on it).\footnote{See e.g., reference \cite{KongCondensable} for further discussion on Lagrangian algebras and their relationship to gapped boundary conditions.}
More specifically, we have
\begin{equation}
    \mathcal{A} = \sum_{\mu \in \mathcal{Z}(\mathcal{C})} V^\mu \cdot \mu,
\end{equation}
where the vector space $V^\mu$ is the space of topological junction operators between the anyon line $\mu$ and the Dirichlet boundary.\footnote{Formally, we have $V^\mu = \mathrm{Hom}_{\mathcal{C}}\left(F(\mu), 1_{\mathcal{C}}\right)$, where $F : \mathcal{Z}(\mathcal{C}) \to \mathcal{C}$ is the forgetful functor. In the notation of \cite{Lin:2022dhv}, we have $V^\mu = W^\mu_1$.} An anyon $\mu$ describes a charge that can carried by a local (not disorder) operator if and only if $\mu \in \mathcal{A}$, i.e., we have $\mathrm{dim}(V^\mu) > 0$. The algebra $\mathcal{A}$ is equipped with a multiplication map $m : \mathcal{A} \otimes \mathcal{A} \to \mathcal{A}$, defined by fusing anyon lines attached to the boundary. In components, the multiplication map is defined by a family of maps
\begin{equation}
    \left(m_{\mu \nu}^{\rho}\right)_i : V^{\mu} \otimes V^{\nu} \to V^{\rho},
\end{equation}
where $i = 1, \dots, N_{\mu \nu}^{\rho}$ runs over the distinct fusion channels $i : \mu \otimes \nu \to \rho$ (see Figure \ref{fig:condensable_algebra_structure}).

We can now describe the set of charges that can acquire a torus one-point function, both with and without additional topological line insertions.

\begin{itemize}
    \item The set of charges $\mu \in \mathcal{Z}(\mathcal{C})$ of local or disorder operators that can acquire a torus one-point function, possibly with other topological line insertions, are those that appear in fusions $\rho \otimes \overline{\rho}$. This set of charges generates the adjoint subcategory $\mathcal{Z}(\mathcal{C})_{\mathrm{ad}}$ of the Drinfeld center. Note that this set of charges is independent of the choice of boundary condition, and is manifestly invariant under any possible orbifolding.

    \item The set of charges $\mu \in \mathcal{A}$ of local operators that can acquire a torus one-point function in the vacuum can be described as follows. First, according to the previous statement, there must exist an anyon $\rho$ such that $\mu$ appears in the fusion $\rho \otimes \overline{\rho}$. For any fusion channel $i : \rho \otimes \overline{\rho} \to \mu$, we compute the state
    \begin{equation}\label{eq:frozen_adjoint}
        \left(m_{\rho \overline{\rho}}^\mu\right)_i \left(\sum_{k} | \psi_k \rangle \otimes | \overline{\psi}_k\rangle \right) \in V^\mu,
    \end{equation}
    where $|\psi_k\rangle$ denotes an orthonormal basis for $V^\rho$. Then a local operator with charge $\mu$ can get only get a nonzero torus one-point function if the state \eqref{eq:frozen_adjoint} in $V^\mu$ is nonzero. This answer is simply an abstract version of the operation illustrated above in Figure \ref{fig:torus_from_sewing}, of sewing together sphere three-point functions by summing over a basis of local operators. The difference is that we have stripped off the physical boundary condition $\mathcal{B}$, obtaining a universal characterization for any 2D CFT with $\mathcal{C}$ symmetry.
\end{itemize}

We strongly suspect that the set of anyons described by \eqref{eq:frozen_adjoint} are precisely those anyons uncharged under the maximal invertible sub-symmetry $\mathcal{C}^\times$, i.e., the group of invertible objects of $\mathcal{C}$.\footnote{Formally, note that the invertible action of $\mathcal{C}^\times$ on topological junctions with the Dirichlet boundary defines a $\mathcal{C}^\times$ action on $\mathcal{A}$, i.e, a functor $B\mathcal{C}^\times \to \mathcal{Z}(\mathcal{C})$, whose value on the basepoint of $B\mathcal{C}^\times$ is given by $\mathcal{A}$. The desired statement is that the $\mathcal{C}^\times$ invariants in $\mathcal{A}$ generate the same fusion subcategory of $\mathcal{Z}(\mathcal{C})$ as the objects characterized by the map \eqref{eq:frozen_adjoint} being nonzero.} One direction is obvious: any anyon that gets a torus one-point function in the vacuum cannot carry charge under any invertible symmetry. While we were unable to provide a proof of the converse, this equivalence holds in every example we considered, including arbitrary modular symmetry categories (Appendix \ref{sec:DIAGONAL_RCFT}) and exotic examples such as Haagerup symmetry (Appendix \ref{sec:HAAGERUP}).

Let us now derive this characterization using the tools of TQFT. Recall that the Turaev-Viro theory (equipped with Dirichlet boundary condition) is a fully extended TQFT with boundary (see e.g., \cite{Lawrence:1993}):
\begin{equation}
    \mathrm{TV}_{\mathcal{C}} : \text{3-Cob}_\partial \to \text{3-Vect},
\end{equation}
characterized through the Cobordism Hypothesis \cite{Baez:1995xq, lurie2009classification} by its value on the point,
\begin{equation}
    \mathrm{TV}_{\mathcal{C}}(\mathrm{pt}) = \mathcal{C}\text{-Mod},
\end{equation}
the 3-vector space of $\mathcal{C}$-module categories, and its value on the half-open interval $\mathbb{I}_* = [0, 1)$ (viewed as a cobordism $\varnothing \to \mathrm{pt}$)
\begin{equation}
    \mathrm{TV}_{\mathcal{C}}\left(\mathbb{I}_*\right) = \mathcal{C},
\end{equation}
viewed as a module category over itself. The category $\mathcal{Z}(\mathcal{C})$ of bulk anyons is given by the value $\mathrm{TV}_{\mathcal{C}}\left(\mathbb{S}^1\right)$ on the circle, and the algebra $\mathcal{A}$ of anyons condensable on the boundary is given by the value $\mathrm{TV}_{\mathcal{C}}\left(\mathrm{Ann}\right)$ on the half-open annulus $\mathrm{Ann} = \mathbb{S}^1 \times \mathbb{I}_* : \varnothing \to \mathbb{S}^1$.

\begin{figure}[t!]
\begin{center}
\includegraphics[scale = 0.75, trim = {1.5cm 6.0cm 0.0cm 8.0cm}]{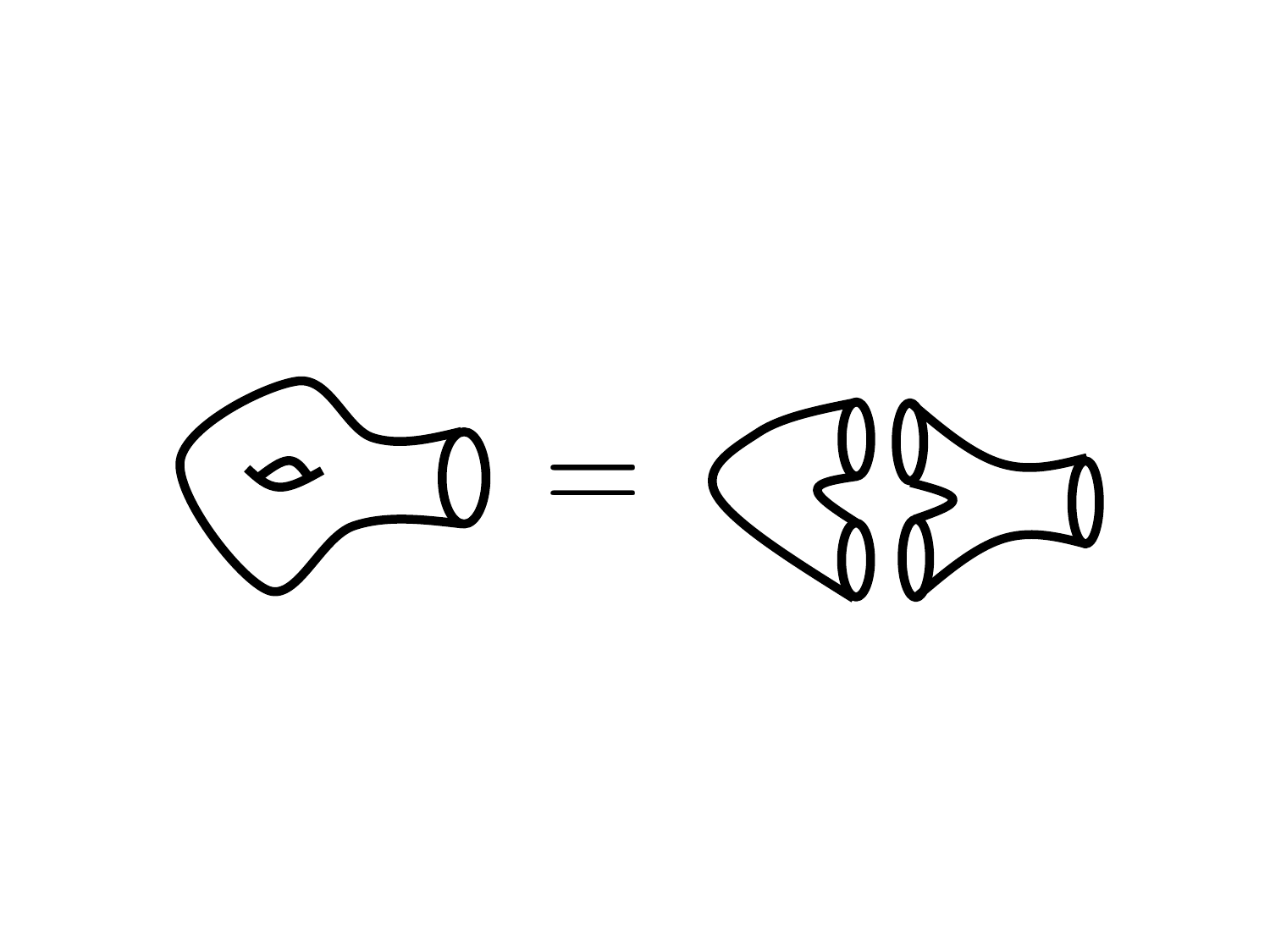}
\caption{We can decompose the punctured torus $\mathbb{T}^2_* : \varnothing \to \mathbb{S}^1$ as the composition of a bent cylinder $\mathrm{Cyl}:\varnothing \to \mathbb{S}^1 \sqcup \mathbb{S}^1$ with a pair-of-pants $\mathrm{Pants}:\mathbb{S}^1 \sqcup \mathbb{S}^1 \to \mathbb{S}^1$.}
\label{fig:torus_as_composition}
\end{center}
\end{figure}

The set of anyons that can acquire a torus one-point function with possible insertions of arbitrary topological lines along the Dirichlet boundary are equivalent to the set of anyons which admit a nonzero state in the defect Hilbert space of the Turaev-Viro theory on the torus with one anyon insertion. To see this, note that any state in this Hilbert space can be prepared by adding arbitrary topological line insertions along the Dirichlet boundary. This defect Hilbert space is characterized by the value $\mathrm{TV}_{\mathcal{C}}\left(\mathbb{T}^2_*\right)$ on a punctured torus. To compute $\mathrm{TV}_{\mathcal{C}}\left(\mathbb{T}^2_*\right)$, note that the punctured torus can be decomposed (see Figure \ref{fig:torus_as_composition}) as a composition
\begin{equation}\label{eq:torus_as_composition}
   \begin{tikzcd}[column sep=huge]
    \varnothing \arrow[r, "\mathrm{Cyl}"] \arrow[dr, "\mathbb{T}^2_*"'] & \mathbb{S}^1 \sqcup \mathbb{S}^1 \arrow[d, "\mathrm{Pants}"] \\
    & \mathbb{S}^1
    \end{tikzcd}
\end{equation}
of two cobordisms: a bent cylinder $\mathrm{Cyl} : \varnothing \to \mathbb{S}^1 \sqcup \mathbb{S}^1$ and a pair-of-pants $\mathrm{Pants} : \mathbb{S}^1 \sqcup \mathbb{S}^1 \to \mathbb{S}^1$. The value of $\mathrm{TV}_{\mathcal{C}}$ on $\mathrm{Cyl}$ and $\mathrm{Pants}$ can be easily computed, so by applying $\mathrm{TV}_{\mathcal{C}}$ to \eqref{eq:torus_as_composition}, we get a commutative diagram in $2$-Vect:
\begin{equation}
   \begin{tikzcd}[column sep=huge]
    \mathrm{Vect} \arrow[r, "\bigoplus_\rho \rho\, \boxtimes\, \overline{\rho}"] \arrow[dr, "\mathrm{TV}_{\mathcal{C}}(\mathbb{T}^2_*)"'] & \mathcal{Z}(\mathcal{C}) \boxtimes \mathcal{Z}(\mathcal{C}) \arrow[d, "\otimes"] \\
    & \mathcal{Z}(\mathcal{C})
    \end{tikzcd}
\end{equation}
Thus, we have $\mathrm{TV}_{\mathcal{C}}(\mathbb{T}^2_*) = \bigoplus_\rho \rho\, \otimes\, \overline{\rho}$, and any anyon $\mu$ that appears in a fusion $\rho \otimes \overline{\rho}$ has a nonzero state in the associated $\mathbb{T}^2$ defect Hilbert space, as claimed.

\begin{figure}[t!]
\begin{center}
\includegraphics[scale = 0.75, trim = {3.0cm 4.0cm 0.0cm 7.0cm}]{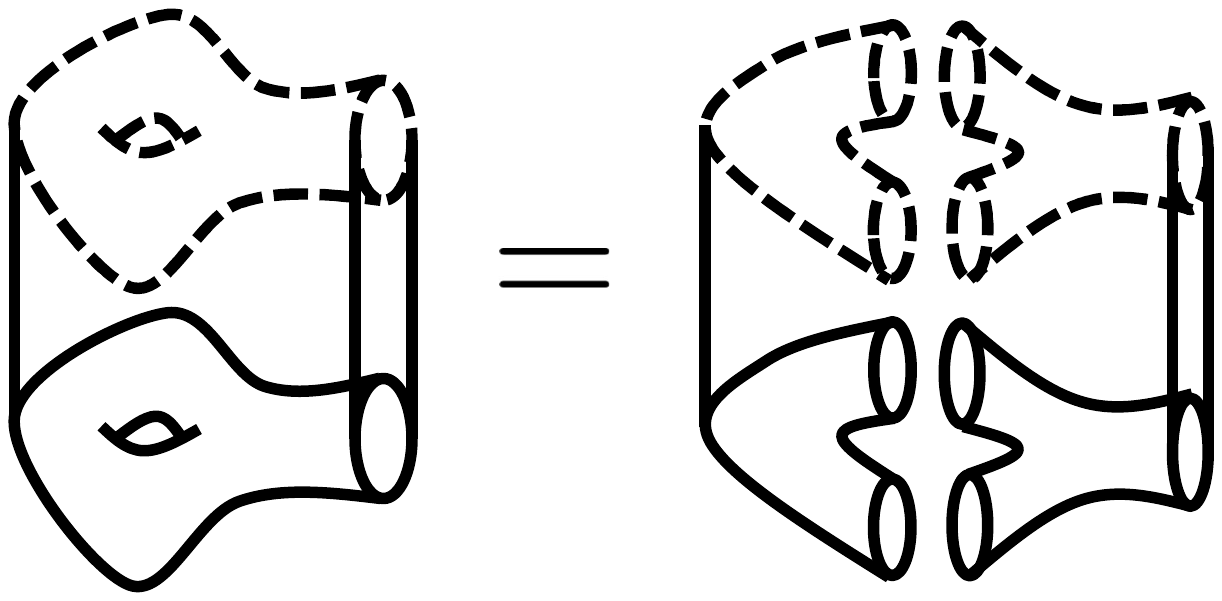}
\caption{The product $\mathbb{T}^2_* \times \mathbb{I}_*$ of the punctured torus with a half-open interval can be decomposed as a composition similarly to how we decomposed the punctured torus itself (Figure \ref{fig:torus_as_composition}). Each manifold with corners pictured here is viewed as a 2-morphism in $(1, 2, 3)$-Cob$_\partial$ from the composition of its left edge with its top edge to its right edge. Note that the bottom edges, given by the Dirichlet boundary, does not count as a source or target from the perspective of $(1, 2, 3)$-Cob$_\partial$. The corresponding commutative diagram in $(1, 2, 3)$-Cob is given by \eqref{eq:321_Cob_diagram}.}
\label{fig:torus_boundary_as_composition}
\end{center}
\end{figure}

Now, the set of anyons that can acquire a torus one-point function in the CFT vacuum is characterized by the boundary state associated to the Dirichlet boundary condition in the punctured torus Hilbert space. This boundary state is described by the value $\mathrm{TV}_{\mathcal{C}}\left(\mathbb{T}^2_* \times \mathbb{I_*} \right)$ on the manifold with corners $\mathbb{T}^2_* \times \mathbb{I_*}$, viewed as a cobordism $\mathbb{T}^2_* \to \mathbb{S}^1 \times \mathbb{I}_*$ from the punctured torus to the annulus (see Figure \ref{fig:torus_boundary_as_composition}). As before, we can view our cobordism as a composition of a bent cylinder and a pair of pants, now both multiplied by $\mathbb{I}_*$. Applying $\mathrm{TV}_{\mathcal{C}}$, we obtain a commutative diagram in the 2-category $(1, 2, 3)$-Cob$_\partial$:
\begin{equation}\label{eq:321_Cob_diagram}
   \begin{tikzcd}[column sep=huge]
    \varnothing \arrow[rr, "\mathrm{Ann}\, \sqcup\, \mathrm{Ann}"'] \arrow[rr, bend left=50, "\mathrm{Cyl}", ""{name=U, below}, start anchor=north east, end anchor=north west] \arrow[rrr,bend right=40, ""{name=D2},"\mathrm{Ann}"'] \ar[rr, bend left=5, phantom, ""{name=D}] \ar[rrr, bend right = 5, phantom, ""{name=U2}] & & \mathbb{S}^1\, \sqcup\, \mathbb{S}^1 \arrow[r, "\mathrm{Pants}"] & \mathbb{S}^1
    \arrow[Rightarrow, from=U, to=D]
    \arrow[Rightarrow, from=U2, to=D2]
    \end{tikzcd}
\end{equation}
Applying $\mathrm{TV}_{\mathcal{C}}$, we get a commutative diagram in $2$-Vect:
\begin{equation}
    \begin{tikzcd}[column sep=huge]
    \mathrm{Vect} \arrow[rr, "\mathcal{A} \, \boxtimes\, \mathcal{A}"'] \arrow[rr, bend left=50, "\bigoplus_\rho \rho\, \boxtimes\, \overline{\rho}", ""{name=U, below}, start anchor=north east, end anchor=north west] \arrow[rrr,bend right=40, ""{name=D2},"\mathcal{A}"'] \ar[rr, bend left=5, phantom, ""{name=D}] \ar[rrr, bend right = 5, phantom, ""{name=U2}] & & \mathcal{Z}(\mathcal{C}) \boxtimes \mathcal{Z}(\mathcal{C}) \arrow[r, "\otimes"] & \mathcal{Z}(\mathcal{C})
    \arrow[Rightarrow, from=U, to=D]
    \arrow[Rightarrow, from=U2, to=D2, "m"]
    \end{tikzcd}
\end{equation}
The top 2-morphism is the map
\begin{equation}
    \bigoplus_\rho \rho \boxtimes \overline{\rho} \to \mathcal{A} \boxtimes \mathcal{A},
\end{equation}
defined on a component $\rho \boxtimes \overline{\rho}$ by the state
\begin{equation}
    \sum_{k} \ket{ \psi_k } \otimes \ket{ \overline{\psi}_k} \in V^\rho \otimes V^{\overline{\rho}},
\end{equation}
where $| \psi_k \rangle$ denotes an orthonormal basis for $V^\rho$ as above. Horizontally composing with the 1-morphism $\mathcal{Z}(\mathcal{C}) \boxtimes \mathcal{Z}(\mathcal{C}) \xrightarrow{\otimes} \mathcal{Z}(\mathcal{C})$ and then vertically composing with the 2-morphism $\mathcal{A} \otimes \mathcal{A} \xrightarrow{m} \mathcal{A}$, we obtain the desired 2-morphism
\begin{equation}\label{eq:TV_frozen_adjoint}
    \mathrm{TV}_{\mathcal{C}}\left(\mathbb{T}^2_* \times \mathbb{I_*} \right) : \bigoplus_{\rho} \rho \otimes \overline{\rho} \to \mathcal{A},
\end{equation}
which characterizes the set of charges $\mu \in \mathcal{A}$ of local operators in the 2D CFT that can acquire a torus one-point function in the vacuum. Expanding \eqref{eq:TV_frozen_adjoint} in components, we recover the desired formula \eqref{eq:frozen_adjoint}.

\subsection{Diagonal RCFTs and Modular Symmetry Categories}\label{sec:DIAGONAL_RCFT}

As an application, we can prove that all non-invertible symmetries are broken by string loops for the special case when the symmetry category $\mathcal{C}$ of our 2D CFT is itself a modular tensor category before passing to the Drinfeld center $\mathcal{Z}(\mathcal{C})$. This includes, in particular, the collection of Verlinde lines of any diagonal RCFT.

Suppose, then, we have a 2D CFT with a modular symmetry category $\mathcal{C}$.\footnote{Note that $\mathcal{C}$ might not describe all the symmetries of our CFT (and certainly does not describe any non-abelian symmetries, since $\mathcal{C}$ is assumed to be braided). For example, consider a WZW model based on a compact Lie group $G$. While the set of Verlinde lines describes all the symmetries that commute with the current algebra, it only includes the invertible symmetries corresponding to the center $Z$ of $G$, and not the larger $(G_L \times G_R)/Z$ invertible symmetry of the WZW model.} By modularity, we have $\mathcal{Z}(\mathcal{C}) = \mathcal{C} \boxtimes \overline{\mathcal{C}}$ as modular tensor categories.\footnote{Here, $\overline{\mathcal{C}}$ denotes the category with the opposite braiding.} The algebra $\mathcal{A}$ associated to the Dirichlet boundary condition of the Turaev-Viro theory $\mathrm{TV}_{\mathcal{C}}$ is given by
\begin{equation}
    \mathcal{A} = \bigoplus_\rho \rho \boxtimes \overline{\rho}.
\end{equation}
Note that all the vector spaces $V^{\rho\, \boxtimes\, \overline{\rho}}$ are one-dimensional. The multiplication map $m : \mathcal{A} \otimes \mathcal{A} \to \mathcal{A}$ is defined via diagonal fusion. In components, for a fusion channel
\begin{equation}
    (i, \overline{\jmath}) : \left( \mu \boxtimes \overline{\mu} \right) \otimes \left( \nu \boxtimes \overline{\nu} \right) \to \rho \boxtimes \overline \rho,
\end{equation}
we have
\begin{equation}
    \left(m_{\mu \nu}^{\rho}\right)_{(i, \overline{\jmath})} = \delta_{i j}.
\end{equation}
Since each $V^{\rho\, \boxtimes\, \overline{\rho}}$ is one-dimensional, there is no sum in \eqref{eq:frozen_adjoint}, and so a charge $\mu \otimes \overline{\mu}$ can get a torus one-point function in the vacuum if and only if $\mu$ appears in a fusion $\rho \otimes \overline{\rho}$. This set of $\mu$ generate the adjoint subcategory $\mathcal{C}_\mathrm{ad}$ of $\mathcal{C}$ itself.

Thus, we need to show that having operators with charges in the adjoint subcategory $\mathcal{C}_\mathrm{ad}$ breaks the non-invertible symmetry down to the maximal invertible subsymmetry $\mathcal{C}^\times$. Now, the symmetry preserved once local operators of charges generating $\mathcal{C}_\mathrm{ad}$ acquire nonzero one-point functions is given\cite{McNamara:2021cuo} by the dual $U_{\mathcal{C}}^\vee$ of the universal grading group of $\mathcal{C}$ \cite[Definition 4.14.2]{etingof2016tensor}. The action of invertible lines on charges by linking defines a grading of $\mathcal{C}$ by the group of characters $\left(\mathcal{C}^\times\right)^\vee$, which induces a map $U_{\mathcal{C}} \to \left(\mathcal{C}^\times\right)^\vee$, or dually, a map $\mathcal{C}^\times \to U_{\mathcal{C}}^\vee$. But by modularity of $\mathcal{C}$, this map is an isomorphism \cite[Theorem 6.3]{gelaki2008nilpotent}, so we see that the nonzero torus one-point functions of operators generating $\mathcal{C}_\mathrm{ad}$ precisely breaks $\mathcal{C}$ to its maximal invertible sub-symmetry $\mathcal{C}^\times$.

\section{Example: Haagerup Symmetry}\label{sec:HAAGERUP}

As evidence that the breaking of all non-invertible symmetries by string loops happens generally, let us quickly verify this effect in the case of a truly exotic non-invertible symmetry in 2D: Haagerup symmetry, described by the Haagerup fusion category $\mathcal{H}_3$, with simple objects
\begin{equation}
    1, \quad \alpha, \quad \alpha^2, \quad \rho, \quad \alpha \rho, \quad \alpha^2 \rho,
\end{equation}
and fusion rules specified by
\begin{equation}\label{eq:Haagerup_fusion}
    \alpha^3 = 1, \quad \alpha \rho = \rho \alpha^2, \quad \rho^2 = 1 + \rho + \alpha \rho + \alpha^2 \rho.
\end{equation}
If there were a counterexample to our general story, one might expect it to be something like Haagerup symmetry: a non-invertible, non-abelian symmetry that cannot be obtained from group-like symmetry or Verlinde lines via discrete gaugings. Nevertheless, we will see that our general story holds true: at one-loop, Haagerup symmetry is broken to its $\mathbb{Z}_3$ invertible sub-symmetry generated by $\alpha$. While there is no formal construction of a 2D CFT with Haagerup symmetry (which could be used in a string compactification), recent numerical evidence favors its existence \cite{Huang:2021nvb}.\footnote{In fact, \cite{Huang:2021nvb} suggests that it might even exist in something as ordinary as a $\mathbb{Z}_3$ orbifold of $T^2$.}

What are the selection rules that Haagerup symmetry places on sphere correlation functions of local operators? The Drinfeld center $\mathcal{Z}(\mathcal{H}_3)$ of the Haagerup fusion category is described in \cite[section 8]{Izumi:2001mi}. We will only need to consider two nontrivial anyons, given by $\pi_1$ and $\pi_2$. The Dirichlet boundary condition is specified by the Lagrangian algebra:
\begin{equation}
    \mathcal{A} = 1 \oplus \pi_1 \oplus \left( \mathbb{C}^2 \cdot \pi_2\right).
\end{equation}
It will help to recall the analogy \cite{Evans:2010yr} between Haagerup symmetry and invertible $S_3$ symmetry; note that if we replaced the third equation in \eqref{eq:Haagerup_fusion} with $\rho^2 = 1$, we would recover a presentation of the symmetric group $S_3$. In this analogy, $\pi_1$ is analogous to the sign representation of $S_3$, while $\pi_2$ is analogous to the standard two-dimensional irreducible representation. Thus, it should not be surprising that local operators $\mathcal{O}_{\pi_1}$ of charge $\pi_1$ are uncharged under the invertible $\mathbb{Z}_3$ sub-symmetry, while local operators $\mathcal{O}_{\pi_2}$ of charge $\pi_2$ come in a multiplet $\mathcal{O}_{\pi_2}^\pm$ of two local operators, with charges $\omega^{\pm 1}$ under the $\mathbb{Z}_3$ symmetry, where $\omega$ is a primitive third root of unity. Moreover, the action of $\rho$ on $\mathcal{O}_{\pi_1}$ is by a sign, while the action of $\rho$ on $\mathcal{O}_{\pi_2}^\pm$ produces $\mathcal{O}_{\pi_2}^\mp$ together with a superposition of disorder operators.

Now, what charges can acquire a nonzero torus one-point function? Clearly, any operator charged in $\pi_2$ cannot, since it has nonzero charge under the invertible $\mathbb{Z}_3$ symmetry. What about operators of charge $\pi_1$? If we took the analogy with $S_3$ too seriously, we might guess that they could not, since there might be cancellations in \eqref{eq:frozen_adjoint} similarly to those that appear for $S_3$ symmetry. If this were true, then Haagerup symmetry would be a counterexample to our general expectation, since the non-invertible symmetry $\rho$ would remain unbroken to all orders in the string loop expansion.

Let us check this guess in the 2D TQFT with Haagerup symmetry constructed in \cite{Huang:2021ytb, Huang:2021zvu}.\footnote{Even though the Haagerup symmetry is spontaneously broken in this TQFT, we should still expect its selection rules to hold if we define correlation functions in a direct sum over all of its distinct vacua, analogously to the selection rules for $S_3$ in the $S_3$ symmetry-breaking TQFT.} This TQFT has six (topological) local operators $1, v, u_1, \overline{u}_1, u_2, \overline{u}_2$. The operator $v$ has charge $\pi_1$, while the operators $u_i, \overline{u}_i$ have charge $\pi_2$. We will compute the torus one-point function $\langle v \rangle_{\mathbb{T}^2}$ by sewing together sphere three point functions. For this, we need the fusion of the local operators with their conjugates, which are given by
\begin{equation}\label{eq:Haagerup_adjoint}
    v \times v = 1 + 3 v, \quad u_1 \times \overline{u}_1 = 1 - \zeta^{-1} v, \quad u_2 \times \overline{u}_2 = 1 + \zeta v,
\end{equation}
where $\zeta = (3 + \sqrt{13})/2 = \langle \rho \rangle$. We can now compute
\begin{equation}\label{eq:Haagerup_one_point}
    \langle v \rangle_{\mathbb{T}^2} = \sum_{\mathcal{O}} \langle v \mathcal{O} \overline{\mathcal{O}} \rangle_{\mathbb{S}^2} = 3 - 2 \zeta^{-1} + 2 \zeta = 9.
\end{equation}
So the torus one-point function $\langle v \rangle_{\mathbb{T}^2}$ is nonzero! The cancellation that would have happened for $S_3$ symmetry does not occur.\footnote{We could have done an analogous calculation for $S_3$; the result can be obtained by taking $3 \to 0, \zeta \to 1$ in \eqref{eq:Haagerup_adjoint} and \eqref{eq:Haagerup_one_point}, reproducing the cancellation in $\langle v \rangle_{\mathbb{T}^2}$.} As a result, we see that the selection rules for Haagerup symmetry are violated at one loop, and even something as exotic as $\mathcal{H}_3$ symmetry on the worldsheet would be broken to its maximal invertible sub-symmetry after considering string loops.

\section{Emergent Non-Invertible Symmetries Beyond  Perturbative Strings} \label{app:break_field}

One of the main points of this paper is that selection rules of non-invertible worldsheet symmetries are generically broken by loop effects in string perturbation theory. When the states running in the loop survive the field theory limit, this string loop contribution can be rewritten as a sum of one-loop field theory diagrams,  which suggests that these sorts of emergent symmetries may also appear in QFT or in quantum gravity away from the perturbative string limit. To illustrate this point, consider a perturbative string background with the Klein bottle as target space.  Since the Klein bottle is a toroidal orbifold
\begin{equation}
    T^2/\mathbb{Z}_2, \quad (X, Y) \sim (X + \pi R, -Y),
\end{equation}
this sigma model has a non-invertible symmetry of the kind discussed in section \ref{sec:continuous_noninvertible}.

We can see echoes of this non-invertible symmetry in the low-energy approximation of string theory (supergravity), even if we keep the Klein bottle large in string units. Reducing any supergravity theory on the Klein bottle and keeping the full Kaluza-Klein (KK) tower, the interaction terms in the supergravity theory lead to couplings between KK modes that respect the selection rule for the non-invertible symmetry: i.e. if we label the KK modes by pairs $(k_X, k_Y)$ of momenta, only defined up to $k_Y \to - k_Y$, interaction terms are only possible if
\begin{equation} \sum_i k_{X, i} = 0, \quad \sum_ i \pm k_{Y, i}=0,\label{pot}\end{equation}
for some choice of $\pm$ signs. This symmetry is likely the low-energy EFT avatar of the one we found in the worldsheet; the fact that the vertices satisfy the selection rules matches with the fact that the non-invertible symmetry is satisfied at tree level. Note that keeping the entire KK tower does not make sense as an EFT in the lower dimensional sense, since the lower-dimensional EFT cutoff is the KK scale.  Instead, it is better understood as a way to organize the higher dimensional EFT when placed on a large Klein bottle; in any case, the KK states exist as physical excitations, and they are long-live enough that one can ask questions about their dynamics.

Breaking of a selection rule like \eqref{pot} at the quantum level is very natural; for instance, an external particle with $Y$ momentum $k_{Y, 1}$ can become a loop pair with momenta $k_{Y, 2}$ and $k_{Y, 1}-k_{Y, 2}$ (respecting \eqref{pot}), and then recombine to form a particle with momentum $-k_{Y, 2}+k_{Y, 1}-k_{Y, 2}=k_{Y, 1}-2k_{Y, 2}$ (analogously to the process illustrated in Figure \ref{fig:torusflow}). The loop then mediates a transition of a particle with $Y$ momentum $k_{Y, 1}$ to another with momentum $k_{Y, 1}-2k_{Y, 2}$, and since $k_{Y, 2}$ was arbitrary, the resulting process manifestly violates \eqref{pot}. In this way ordinary, field theory loop effects can violate a non-invertible symmetry preserved by the couplings, just like in string theory. Again, we emphasize that that the non-invertible symmetry does not act in any EFT of a fixed dimension: the states we are considering are above the cutoff of the lower-dimensional EFT, while the non-invertibility is coming from the Klein bottle reduction, and would not appear in the fully decompactified theory.

Because the symmetry is broken by quantum effects, one expects it to become a good approximate close to any classical limit, even away from the perturbative string. For instance, we could consider M-theory on the Klein bottle. Its low-dimensional expansion, 11D supergravity, is a power series expansion in powers of the 11D Planck mass $M_{11}$;  in the decompactification limit, when the characteristic size of the Klein bottle $R$ is very large in 11D Planck units,  quantum corrections that break the non-invertible symmetry are naturally suppressed in powers of $1/(M_{11}R)$, and they should vanish in the limit, where the non-invertible symmetry becomes a subsymmetry of higher-dimensional Poincar\'{e}, which are exact. It would be interesting to check this example in detail; although we have not done so, we have checked that corrections are suppressed in this way for a simple toy model (a $\Phi^3$ theory) when a regularization preserving higher-dimensional Poincar\'{e} symmetry is used, as in \cite{Cheng:2002iz,Martinez-Pascual:2020cbv}.

From this point of view, it may be that the phenomenon of non-invertible selection rules of the classical action being weakly broken by quantum effects is not necessarily an intrinsically stringy phenomenon, but rather also appears as a general feature of compactification on manifolds with local isometries that fail to be well-defined globally. Clearly, it would be interesting to flesh this story out in more detail.


\bibliographystyle{utphys}
\bibliography{NonInvQG}

\end{document}